\documentclass[lettersize, final, twoside, journal]{IEEEtran}
\usepackage{cite}
\usepackage{amsmath, amssymb, amsfonts}
\usepackage{enumitem}
\usepackage{algorithm}
\usepackage{algorithmic}
\usepackage{graphicx}
\usepackage{subfigure}
\usepackage{textcomp}
\usepackage{xcolor}
\usepackage{amsthm}
\usepackage{multirow}
\usepackage{bm}
\usepackage{siunitx}
\usepackage[]{hyperref} 
\hypersetup{
hidelinks,
colorlinks=false,
linkcolor=black,
citecolor=black,
}
\allowdisplaybreaks
\usepackage{balance}
\usepackage{longtable, threeparttable}
\usepackage{makecell}

\newtheorem{theorem}{Theorem}

\begin{document}

\title{Energy-Efficient Index and Code Index Modulations for Spread CPM Signals in Internet of Things}

\author{Long Yuan, Wenkun Wen,~\IEEEmembership{Member,~IEEE}, Junlin Liu, Peiran Wu,~\IEEEmembership{Member,~IEEE}, \\ and Minghua Xia,~\IEEEmembership{Senior Member,~IEEE}

	\thanks{Manuscript received 10 July 2025; revised 31 July 2025; accepted 09 August  2025. \textit{(Corresponding authors: Wenkun Wen; Minghua Xia.)}

	Long Yuan,  Peiran Wu, and Minghua Xia are with the School of Electronics and Information Technology, Sun Yat-sen University, Guangzhou 510006, China (email: yuanlong@mail2.sysu.edu.cn, wupr3@mail.sysu.edu.cn, xiamingh@mail.sysu.edu.cn)
	
	Wenkun Wen and Junlin Liu are with the R\&D Department of the Techphant Technologies Co. Ltd., Guangzhou 510310, China (email: wenwenkun@techphant.net, liujunlin@techphant.net)

	Color versions of one or more of the figures in this article are available online at https://ieeexplore.ieee.org.
	
	Digital Object Identifier 

	Copyright (c) 2025 IEEE. Personal use of this material is permitted. However, permission to use this material for any other purposes must be obtained from the IEEE by sending a request to pubs-permissions@ieee.org.
}
}

\markboth{IEEE Internet of Things Journal} {Yuan \MakeLowercase{\textit{et al.}}: Energy-Efficient Index and Code Index Modulations for Spread CPM Signals in Internet of Things}

\maketitle

\IEEEpubid{\begin{minipage}{\textwidth} \ \\[12pt] \centering 2327-4662 \copyright\ 2025 IEEE. All rights reserved, including rights for text and data mining, and training of artificial intelligence \\ and similar technologies. Personal use is permitted, but republication/redistribution requires IEEE permission. \\
See \url{https://www.ieee.org/publications/rights/index.html} for more information.\end{minipage}}

 \IEEEpubidadjcol

\begin{abstract}
	The evolution of Internet of Things technologies is driven by four key demands: ultra-low power consumption, high spectral efficiency, reduced implementation cost, and support for massive connectivity. To address these challenges, this paper proposes two novel modulation schemes that integrate continuous phase modulation (CPM) with spread spectrum (SS) techniques. We begin by establishing the quasi-orthogonality properties of CPM-SS sequences. The first scheme, termed IM-CPM-SS, employs index modulation (IM) to select spreading sequences from the CPM-SS set, thereby improving spectral efficiency while maintaining the constant-envelope property. The second scheme, referred to as CIM-CPM-SS, introduces code index modulation (CIM), which partitions the input bits such that one subset is mapped to phase-shift keying symbols and the other to CPM-SS sequence indices. Both schemes are applied to downlink non-orthogonal multiple access (NOMA) systems. We analyze their performance in terms of bit error rate (BER), spectral and energy efficiency, computational complexity, and peak-to-average power ratio characteristics under nonlinear amplifier conditions. Simulation results demonstrate that both schemes outperform conventional approaches in BER while preserving the benefits of constant-envelope, continuous-phase signaling. Furthermore, they achieve higher spectral and energy efficiency and exhibit strong resilience to nonlinear distortions in downlink NOMA scenarios.
\end{abstract}

\begin{IEEEkeywords}
	Code index modulation, continuous phase modulation, index modulation, Internet of Things, non-orthogonal multiple access, spread spectrum. 
\end{IEEEkeywords}

\section{Introduction}
As a cornerstone of the next-generation information infrastructure, the Internet of Things (IoT) connects various end devices to the Internet, enabling seamless interaction between devices and between people and devices. This interconnectivity facilitates data collection, interaction, and sharing, making people's daily lives more convenient and efficient. However, conventional cellular systems such as 4G and 5G fall short in addressing the unique needs of IoT, including high deployment costs, significant energy consumption, and scalable coverage. In response, low-power wide-area network (LPWAN) technology has been introduced to meet the IoT's requirements for long-range coverage, low power consumption, and cost-effectiveness \cite{LPWA}. Several representative LPWAN technologies have emerged, including NB-IoT, which operates in licensed spectrum \cite{NB-IoT-2025}, and LoRa and Sigfox, which use unlicensed bands \cite{LoRa, Sigfox}. Standards like IEEE 802.15.4 also contribute to the development of low-rate wireless networks \cite{2024-IEEE-standard}. 

To meet the long-range communication demands of LPWANs, spread spectrum (SS) techniques are frequently adopted. SS enhances receiver sensitivity and improves resistance to interference and multipath fading, as evidenced by its use in cellular, military, underwater, and secure communication systems, dating back to the release of Interim Standard 95 (IS-95) in 1995. Meanwhile, advanced physical layer technologies lie at the foundation of IoT communications. Continuous phase modulation (CPM) signals, characterized by their constant envelope, continuous phase, compact spectrum, and compatibility with nonlinear power amplifiers, allow for effective use of power and spectrum \cite{CPM-1981}. These attributes make CPM particularly well-suited for wireless applications with strict constraints on power and bandwidth. In particular, CPM is highly applicable to communications in resource-constrained, energy-efficient IoT systems, making it ideal for ``green'' IoT applications \cite{CPM-IoT-24}. Consequently, combining CPM with SS techniques, referred to as CPM-SS, offers a compelling solution for the technical requirements of LPWAN \cite{CPM-DSSS}.

 \IEEEpubidadjcol
 
Despite its advantages, traditional CPM-SS receivers typically perform demodulation and despreading as separate processes, resulting in higher complexity and performance degradation \cite{CPM-DSSS-GNSS}. Recent work has focused on joint demodulation and despreading to improve receiver performance \cite{Iot-zhangruiqi, Wen25IoTJ}. However, many conventional schemes still assign one bit to each spreading sequence, which limits the spectral efficiency. Some methods have introduced multiple spreading codes to enhance the system's spectral efficiency, but they often increase complexity and do not fully resolve performance tradeoffs \cite{DSSS-CPM}. This highlights the need for a solution that combines the benefits of both SS and CPM, while increasing the spectral efficiency and maintaining manageable complexity. Such a technology would significantly improve overall system performance, ensuring greater efficiency and capability in meeting the evolving demands of scalable IoT systems.

In recent years, index modulation (IM) has emerged with advantages in energy and spectral efficiency. These characteristics make it promising for next-generation communication technologies \cite{IM-NextG, SM-Survey}. IM divides the information bits to be transmitted into two parts. One part serves as an index and controls the activation of communication resources, such as time slots, subcarriers, antennas, or spreading codes. In contrast, the other part is used for traditional modulation techniques, such as phase shift keying (PSK) and quadrature amplitude modulation \cite{Multidim-IM}. The modulation symbols are transmitted through the resources activated by the index. Compared to traditional modulation methods, since the index portion consumes no additional transmit power, IM enables improved spectral efficiency without increased energy usage.

Several IM-based techniques have outperformed conventional approaches. For example, spatial modulation (SM) surpasses multi-input multi-output (MIMO) systems \cite{SM}, orthogonal frequency division multiplexing index modulation (OFDM-IM) outperforms conventional OFDM systems \cite{OFDM-IM}, and code index modulation (CIM) exceeds direct sequence spread spectrum (DSSS) systems \cite{CIM}. Among them, CIM combines the benefits of both SS and IM, making it especially attractive for LPWAN applications. CIM has been extensively studied in recent years. For example, CIM has been applied to chaos-based communication systems to enhance performance, particularly in terms of robustness and spectral efficiency \cite{CIM-OM-MDCSK, JCCIM-MDCSK}. Additionally, the recently proposed quadruple code index modulation aims to further improve wireless transmission performance by offering higher spectral efficiency and reduced complexity \cite{QCIM}. However, most current CIM implementations use orthogonal or quasi-orthogonal bipolar spread spectrum codes, which yield unstable signal envelopes. When passed through a nonlinear transmitter, these signals are prone to distortion, leading to significant performance degradation.

At the same time, the exponential growth of IoT devices stresses traditional orthogonal multiple access (OMA) methods, which lack scalability under constrained resources. Non-orthogonal multiple access (NOMA) addresses this by allowing multiple users to share the same time-frequency resources by superimposing their signals at the transmitter \cite{NOOB-2012, NOMA-IoT}. However, signal superposition in downlink NOMA increases envelope variation and leads to a high peak-to-average power ratio (PAPR), making signals more sensitive to nonlinear distortion \cite{CPM-MA}. In real-world applications, input power back-off (IBO) is often employed to mitigate nonlinear distortion, but it comes at the cost of reduced power amplifier efficiency. Consequently, signals that naturally have a low PAPR are preferred, as they enable power amplifiers (PAs) to operate more efficiently with less IBO. CPM in downlink NOMA transmission inherently exhibits low PAPR, making it more suitable for nonlinear hardware and better aligned with the cost and energy constraints of IoT devices \cite{CPM-NOMA}. Despite CPM's potential, the integration of CIM and CPM-SS within downlink NOMA systems remains largely unexplored.

This paper builds on our recent efforts \cite{Long-ICCC, Wen-ICCC} to propose two new modulation schemes, {\bf IM-CPM-SS} and {\bf CIM-CPM-SS}, that combine CPM-SS with IM and CIM, respectively. These schemes leverage the constant-envelope and phase-continuous properties of CPM to develop IoT-oriented solutions that support low power consumption, reduced costs, and improved spectral efficiency. Furthermore, we apply the proposed CPM-SS-based schemes to downlink NOMA systems to achieve lower PAPR, higher spectral efficiency, enhanced energy efficiency, and robustness against nonlinear distortion.

The main contributions of this work are as follows:
\begin{itemize}
	\item We establish the quasi-orthogonality properties of CPM-SS sequences, enabling their applications in efficient spreading and multiple-user access.
	\item We design two modulation schemes, IM-CPM-SS and CIM-CPM-SS, and develop a corresponding downlink NOMA system model that leverages the proposed signal structure for LPWAN applications.
	\item We derive analytical bit error rate (BER) expressions under both additive white Gaussian noise (AWGN) and Rayleigh fading channels, and analyze spectral efficiency, energy efficiency, and computational complexity.
	\item We model a nonlinear power amplifier in the downlink NOMA system to assess signal robustness against nonlinear distortion, comparing PAPR characteristics against existing solutions.
	\item Extensive simulation and analytical results validate the proposed schemes, demonstrating superior BER, reduced PAPR, and improved robustness against nonlinear distortion, making them suitable for practical IoT deployment.
\end{itemize}

The remainder of this paper is organized as follows: Section~\ref{Section-II} demonstrates the quasi-orthogonality of CPM-SS sequences. Section~\ref{Section-III} develops the modulation schemes and the system model for downlink NOMA transmission. Section~\ref{Section-IV} analyzes the performance of the proposed schemes and introduces a nonlinear model to assess the ability of downlink NOMA transmission to resist nonlinear distortion. Section~\ref{Section-V} presents and discusses the simulation and analytical results for the proposed schemes, followed by a conclusion in Section~\ref{Section-VI}. 

{\it Notation:}  The superscripts $(\cdot)^T$ and $(\cdot)^H$ denote the transpose and Hermitian transpose, respectively.  The operators $\Vert \cdot \Vert_2$, $\mathbb{E}(\cdot)$, and ${\Re}(\cdot)$ represent the Euclidean $\ell_2$-norm, mathematical expectation, and the real part of a complex-valued number, respectively. The symbol $j \triangleq \sqrt{-1}$ refers to the imaginary unit, and $\binom{n}{k} \triangleq \frac{n!}{k!(n-k)!}$ denotes the binomial coefficient. The symbol $\text{Pr}(\cdot)$ counts the probability of an event. The special function $I_0(x) \triangleq \sum_{k=0}^{\infty} (k!)^{-2} \left(x/2\right)^{2k}$ refers to the zero-order modified Bessel function of the first kind, and $\text{erfc}(x) \triangleq \frac{2}{\sqrt{\pi}} \int_{x}^{\infty} \exp(-y^2) \, {\rm d}y$ defines the complementary error function.

\section{Mathematical Preliminary: The Quasi-Orthogonality of CPM-SS Sequences} 
\label{Section-II}
This section introduces the spread CPM signals and analyzes the quasi-orthogonality properties of the CPM-SS sequences. These properties play a crucial role in enabling efficient spreading and supporting multiple access in resource-constrained IoT environments.

Given an $M$-ary symbol $a_m \in \{\pm1, \cdots, \pm (M-1)\}$, the phase of a CPM signal can be expressed as \cite[Eq.~(3.3-14)]{Digital-Communication}
\begin{equation}
    \bm{\phi}(t; \bm{a}) = 2\pi h \hspace{-.5em} \sum\limits_{m=-\infty}^{n} \hspace{-.5em} a_m q(t-mT),\ nT\leq t \leq (n+1)T ,
\label{eq-CPM}
\end{equation}
where $h = h_m/h_d$ is the modulation index, with $h_m$ and $h_d$ being relatively prime integers, which determines the amount of phase change corresponding to each symbol; $T$ denotes the symbol interval, and $q(t)$ stands for the phase-smoothing function defined as
\begin{equation} \label{eq-CPM-qt}
q(t) = 
\left\{\begin{array}{rl}
            \int_{0}^{t} g(\tau) \, {\rm d}\tau, & \text{if }  0 \leq t \leq LT;    \\
           1/2,  &  \text{if }  t > LT, 
\end{array} \right.
\end{equation}
where $g(\tau)$ denotes the frequency pulse shape and $L$ is the memory length of the CPM signal. The cases $L = 1$ and $L > 1$ correspond to full-response and partial-response CPM, respectively. Commonly used pulse shapes include rectangular (REC) and raised cosine (RC) for full-response, and spectral raised cosine (SRC) and Gaussian (GAU) for partial-response. Based on \eqref{eq-CPM-qt}, the phase given by \eqref{eq-CPM} can be decomposed as
\begin{equation}
    \bm{\phi}(t; \bm{a})=\underbrace{\pi h \sum\limits_{m=0}^{n-L} a_m}_{\phi_{n-L}} + \underbrace{2\pi h \hspace{-1em} \sum\limits_{m=n-L+1}^{n} \hspace{-1em} a_m q(t-mT)}_{\phi(t; \bm{a}_n)},
\label{eq-CPM-v}
\end{equation}
where $nT \leq t \leq (n+1)T$, and the term $\phi_{n-L}$ represents the accumulated phase at $t = nT$, and $\phi(t; \bm{a}_n)$ denotes the instantaneous phase determined by the most recent $L$ symbols, i.e., $\bm{a}_n = \{a_{n-L+1}, \cdots, a_n\}$. Accordingly, the phase state at time $t = nT$ is represented as 
\begin{equation}
    s_n = \{\phi_{n-L}, a_{n-L+1}, \cdots, a_{n-1}\}.
\label{eq-CPM-phase}
\end{equation}
In general, the number of possible phase states is $h_d M^{L-1}$ if $h_m$ is even, and $2h_d M^{L-1}$ if $h_m$ is odd \cite[Eq.~(3.3-18)]{Digital-Communication}.

We now introduce the CPM-based spread spectrum scheme (CPM-SS), where CPM signals are spread using sequences derived from an $m$-sequence. Let the original codebook be $\bm{X} = [\bm{x}_1, \cdots, \bm{x}_N]^T$, where each codeword $\bm{x}_n = [x_{n,1}, \cdots, x_{n,N}]^T$ is generated by cyclically shifting the $m$-sequence. Here, $x_{n, p} \in \{-1, 1\}$ denotes the bipolar mapping of binary bits for all $n, p = 1, \cdots, N$. The code length is given by $N = 2^{\text{SF}} - 1$, where $\text{SF}$ represents the spreading factor. Each codeword in $\bm{X}$ is modulated using CPM to produce the complex-valued codebook $\bm{Y} = [\bm{y}_1, \cdots, \bm{y}_N]^T$, where each $\bm{y}_n \in \mathbb{C}^{NP \times 1}$ represents the $n^\text{th}$ modulated codeword, with $P$ being the oversampling factor per chip. Notably, the constant-envelope property of the CPM signal is preserved after spreading with the $m$-sequence. This is because the spreading is applied multiplicatively, typically through binary phase modulation (i.e., multiplying by $\pm1$), which alters the phase but does not affect the signal's amplitude. As a result, the instantaneous envelope of the CPM signal remains constant, and the inherent power efficiency of CPM is retained.

To characterize the correlation between codewords, we analyze their cross-correlation, formalized in the theorem below.
\begin{theorem} \label{The-1}
Let $d$ denote the number of cyclic shifts of the original $m$-sequence, with $d > 0$ for a right shift and $d < 0$ for a left shift. Then, as $N \to \infty$, the cross-correlation coefficient $\rho$ between two modulated codewords $\bm{y}_n$ and $\bm{y}_l$ satisfies:
\begin{equation} \label{eq-crosscorr-final}
    \rho = 
    \left\{\begin{array}{rl}
           \pm j\frac{1}{\pi}, & \text{if }  |n-l| = 1; \\
            0, & \text{if } |n-l| >1, 
          \end{array} \right.
\end{equation}
\end{theorem}

\begin{IEEEproof}
See Appendix.
\end{IEEEproof}

By Theorem~\ref{The-1}, the real part of the cross-correlation coefficient remains small enough to enable coherent detection for all codewords in $\bm{Y}$ \cite{LoRa-2019}. However, for noncoherent detection and to facilitate CIM implementation, we further refine the codebook to suppress cross-interference.

Specifically, we construct a subset $\bm{C}$ by discarding codewords that are adjacent in shift order and thus exhibit high cross-correlation (i.e., $\pm j/\pi$). Each selected codeword $\bm{c}_n \in \mathbb{R}^{N \times 1}$ is then modulated using CPM to form a new complex-valued codebook $\bm{Z} = [\bm{z}_1, \cdots, \bm{z}_{N_c}]^T$, where
\begin{equation}
\bm{z}_n = \exp(j \phi(t; \bm{c}_n)) \in \mathbb{C}^{NP \times 1}.
\end{equation}

This codebook refinement guarantees that the maximum magnitude of the cross-correlation coefficient between any two codewords in $\bm{Z}$ is upper-bounded by $1/N$, as demonstrated in the proof of Theorem~\ref{The-1}, thereby implying pairwise quasi-orthogonality in the asymptotic regime as $N \to \infty$. As a result, the total number of usable codewords, denoted by $N_c$, satisfies
\begin{align} \label{eq-Nc}
N_c \leq 2^{\text{SF} - 1} - 1.
\end{align}

\floatname{algorithm}{Algorithm}
\begin{algorithm}[!t]
    \caption{Generating the Codebook $\bm{Z}$}
    \label{alg-generating-Z}
    \small
    \renewcommand{\algorithmicrequire}{\textbf{Input:}}
    \renewcommand{\algorithmicensure}{\textbf{Output:}}
    \begin{algorithmic}[1]
        \REQUIRE $q(t)$, $h$, $L$, $P$, and the original $m$-sequence of length $N$;
        \ENSURE $\bm{Z}$
        \STATE The original $m$-sequence is mapped to the bipolar sequence;
        \STATE The original codebook $\bm{X}_{N \times N}$ is generated by cyclically shifting the bipolar sequence;
        \STATE According to the parameters $q(t)$, $h$, $L$, and $P$, the CPM can be implemented by Eq.~\eqref{eq-CPM};
        \STATE Each codeword $\bm{x}_n$ in $\bm{X}$ is modulated by the CPM to obtain the complex-valued codeword $\bm{y}_n$;
        \FOR{each $n \in [1,N]$}
	        \STATE $\bm{y}_n$ = exp($j \phi(t; \bm{x}_n)$);
        \ENDFOR
        \STATE Let $\bm{Y} = [\bm{y}_1, \cdots, \bm{y}_{N}]^T$;
	\STATE The codebook $\bm{Z}$ is obtained by discarding codewords in $\bm{Y}$ that are adjacent in shift order.
    \end{algorithmic}
\end{algorithm}

For ease of reproducibility, {\bf Algorithm~\ref{alg-generating-Z}} outlines the steps for generating the codebook $\bm{Z}$. This selective codebook construction forms the basis for the proposed IM-CPM-SS and CIM-CPM-SS schemes, enabling robust, spectrally efficient transmission in LPWAN and NOMA settings.

\section{Transceiver Design and Multi-Access Scheme} \label{Section-III}
This section introduces the point-to-point transceiver architectures for the proposed IM-CPM-SS and CIM-CPM-SS modulation schemes, and then extends them to a multi-user downlink NOMA scenario.

\subsection{Point-to-Point Transceiver Design} \label{Section-III-A}
In a conventional CPM-SS system, binary input bits are first spread using a sequence derived from an $m$-sequence and subsequently modulated via CPM. At the receiver, maximum likelihood sequence detection based on the Viterbi algorithm is typically employed to demodulate the CPM signal, followed by a separate despreading stage to recover the original bits. However, this separation introduces two significant limitations: 
\begin{itemize}
	\item Performance degradation due to the lack of joint processing, and
	\item Limited spectral efficiency due to one bit per codeword mapping.
\end{itemize}
To address these issues, we propose two improved modulation schemes, IM-CPM-SS and CIM-CPM-SS, that jointly leverage the quasi-orthogonality of CPM-SS codewords and the resource indexing flexibility of index modulation. These designs improve both detection performance and spectral efficiency while maintaining constant-envelope signaling.

In the {\bf IM-CPM-SS} scheme, each group of input bits is mapped to an index that selects a quasi-orthogonal CPM-SS codeword. This mapping transmits multiple bits per symbol interval without increasing energy consumption.

In the {\bf CIM-CPM-SS} scheme, each bit group is split into two parts: one segment determines the codeword index, and the other is mapped to a PSK symbol. The resulting signal combines the selected CPM-SS codeword and the PSK symbol via element-wise multiplication, offering a tunable trade-off between throughput and reliability.

Both schemes employ joint demodulation and despreading at the receiver to avoid the inefficiencies of separate processing, making them well-suited for energy- and resource-constrained IoT systems.

\begin{figure}[!t]
\centerline{\includegraphics[width=0.50\textwidth]{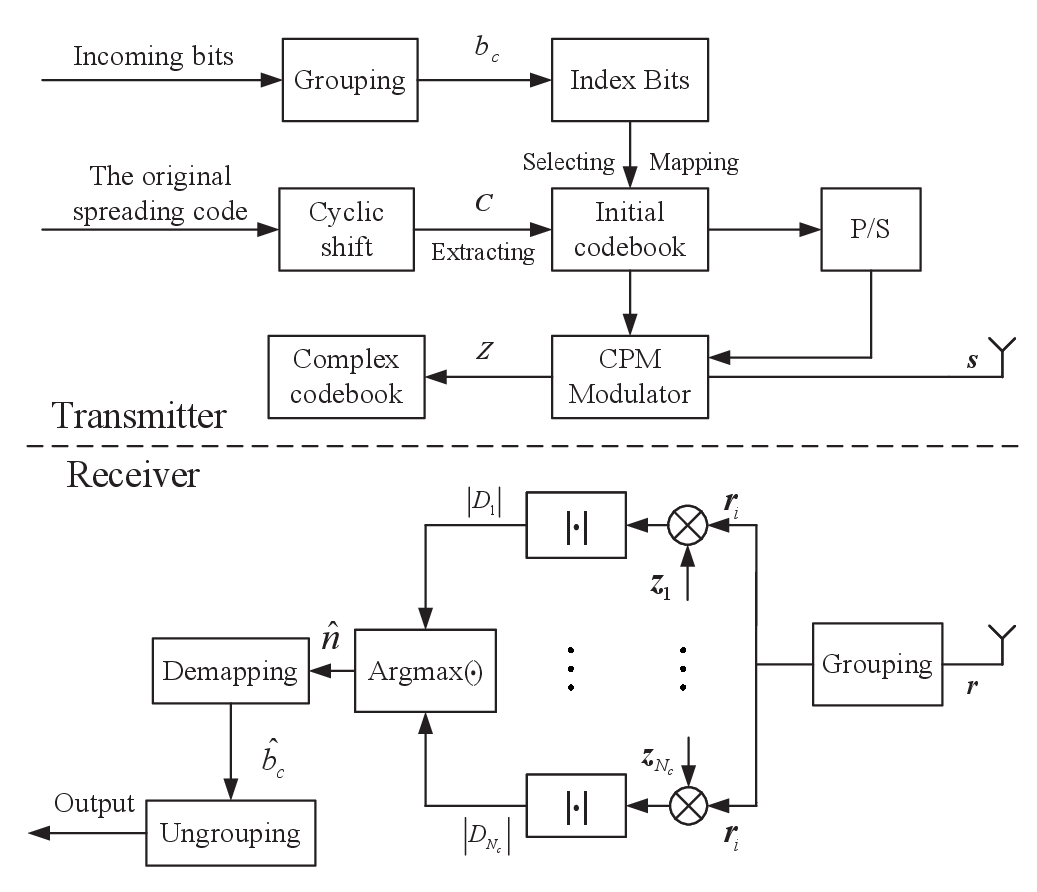}}
\vspace{-10pt}
\caption{Block diagram of the proposed IM-CPM-SS transceiver system.}
\label{fig-IM-CPM-SS-transceiver}
\end{figure}

\subsubsection{\underline{\textbf{IM-CPM-SS Scheme}}}
Fig.~\ref{fig-IM-CPM-SS-transceiver} shows the transceiver architecture of the proposed IM-CPM-SS scheme. The transmitter (upper panel) constructs a codebook $\bm{C}$ by cyclically shifting an $m$-sequence and selecting a subset of non-adjacent, quasi-orthogonal codewords. Each codeword is then modulated using CPM to yield the complex-valued codebook $\bm{Z}$.

The input bitstream is partitioned into $I$ groups of $b_c$ bits each, referred to as index bits. Each group is converted to a decimal index that selects a codeword $\bm{z}_n$ from $\bm{Z}$. The selected codewords are serialized and modulated using CPM for transmission.

At the receiver (lower panel of Fig.~\ref{fig-IM-CPM-SS-transceiver}), assuming perfect synchronization, the received signal $\bm{r}$ is segmented into $I$ groups. For the $i^\text{th}$ group, the received signal is modeled in baseband as
\begin{align}
	\bm{r}_i = h_i \bm{z}_i + \bm{n}_i,
\label{eq-scheme1-r}
\end{align}
where $i = 0, \cdots, I-1$, $\bm{r}_i \in \mathbb{C}^{NP \times 1}$ denotes the received vector, with $P$ being the oversampling factor per chip; $h_i$ is a complex-valued channel coefficient, and $\bm{n}_i$ represents an AWGN vector with zero mean and covariance $N_0 \bm{I}$, with $\bm{I}$ referring to identity matrix of appropriate size.

To detect the transmitted information, the receiver computes the cross-correlation between $\bm{r}_i$ and all codewords in $\bm{Z}$. The transmitted index is estimated as:
\begin{align}
\hat{n} = \mathop{\arg\max}\limits_{n} \left| \bm{r}_i^H \bm{z}_n \right|, \, \forall \bm{z}_n \in \bm{Z}.
\label{eq-scheme1-i}
\end{align}
Finally, the estimated index $\hat{n}$ is converted back to binary, and the recovered groups are concatenated to reconstruct the original bitstream.

\begin{figure}[!t]
\centerline{\includegraphics[width=0.50\textwidth]{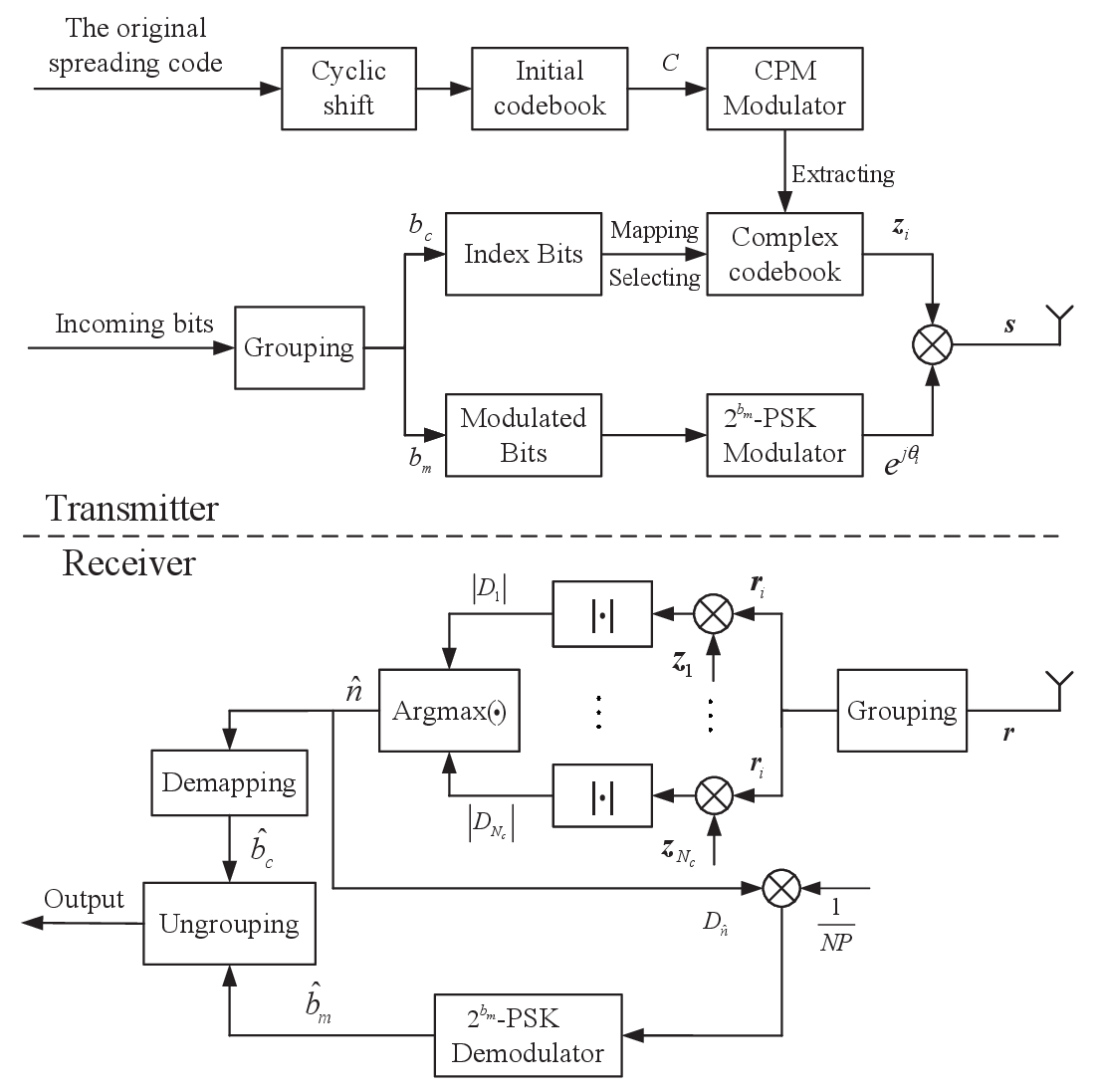}}
\vspace{-10pt}
\caption{Block diagram of the proposed CIM-CPM-SS transceiver system.}
\label{fig-CIM-CPM-SS-transceiver}
\end{figure}

\subsubsection{\underline{\textbf{CIM-CPM-SS Scheme}}}
Fig.~\ref{fig-CIM-CPM-SS-transceiver} illustrates the system model of the proposed CIM-CPM-SS transceiver. At the transmitter side (upper panel), a complex-valued codebook $\bm{Z}$ is first generated using the same procedure described for the IM-CPM-SS scheme, based on CPM-modulated, quasi-orthogonal spreading codewords.

The input bitstream is divided into $I$ groups, each containing $b_g = b_c + b_m$ bits. Here, $b_c$ bits are designated as index bits and $b_m$ bits as modulated bits. The index bits are converted from binary to decimal to select a codeword $\bm{z}_n$ from the complex codebook $\bm{Z}$ for spreading. Simultaneously, the modulated bits are mapped to a symbol from a $2^{b_m}$-ary PSK constellation. The final baseband transmission signal $\bm{s}_i$ for the $i^\text{th}$ group is obtained by element-wise multiplication of the selected spreading codeword and the PSK-modulated symbol.

At the receiver (lower panel), the received signal is segmented into $I$ groups. The $i^\text{th}$ received baseband signal can be modeled as
\begin{align}
\bm{r}_i = h_i \bm{z}_i e^{j\theta_i} + \bm{n}_i,
\label{eq-scheme2-r}
\end{align}
where $h_i$ denotes the complex-valued channel coefficient, $e^{j\theta_i}$ is the transmitted PSK symbol, and $\bm{n}_i$ is zero-mean, circularly symmetric complex AWGN with covariance matrix $N_0 \bm{I}$.

Ideally, a joint maximum likelihood sequence estimation could be employed to recover both the index and modulated bits. However, due to its prohibitively high computational complexity, we propose an alternative, yet efficient, detection algorithm.

Assuming perfect channel state information, i.e., $h_i$ is known at the receiver, we first estimate the index bits by correlating the received signal $\bm{r}_i$ with all codewords in $\bm{Z}$ and selecting the one that yields the maximum correlation magnitude. After compensating for the channel, the received signal can be rewritten as:
\begin{align}
\bm{r}_i = |h_i|^2 \bm{z}_i e^{j\theta_i} + h_i^H \bm{n}_i.
\label{eq-scheme2-channel-compensation}
\end{align}
The correlator output for codeword $\bm{z}_n$ is defined as the decision variable:
\begin{equation}
    D_{i,n} = 
    \left\{\begin{array}{rl}
           |h_i|^2 E_s e^{j \theta_i} + h_i^H \bm{z}_n^H \bm{n}_i, & \text {if } n = \hat{n};  \\
            h_i^H \bm{z}_n^H \bm{n}_i,  & \text{otherwise}, 
    \end{array} \right.
\label{eq-scheme2-detect-variable}
\end{equation}
where $E_s \triangleq \bm{z}_n^H \bm{z}_n$ is the energy of the CPM-SS signal. Then, the estimated index is obtained as:
\begin{align}
\hat{n} = \mathop{\arg\max}\limits_{n} |D_{i,n}|.
\label{eq-scheme2-codeindex}
\end{align}

The detected index $\hat{n}$ is converted back into binary to recover the index bits $\hat{b}_c$. Using $\hat{n}$, the received modulation symbol is estimated by normalizing the correlator output: $D_{i,\hat{n}} / NP$. This symbol is then demodulated using a $2^{b_m}$-ary PSK demodulator to obtain the modulated bits $\hat{b}_m$. Finally, the recovered index bits $\hat{b}_c$ and modulated bits $\hat{b}_m$ are concatenated to reconstruct the original transmitted bitstream.

\subsection{Downlink NOMA Transmission}
\label{Section-III-B}
This subsection extends the point-to-point transceiver architecture to a multi-user downlink scenario using NOMA. Consider a system with $U$ users served simultaneously by a base station. The received signal at user $u$ is given by:
 \begin{align}
	\bm{r}_u = \sum_{q=1}^U h_u \sqrt{P_q} \bm{s}_q + \bm{n}_u, 
\label{eq-NOMA-r}
\end{align}
where $P_q$ denotes the transmit power allocated to user $q$, $\bm{s}_q$ is the transmit signal intended for user $q$, $h_u$ represents the complex-valued channel coefficient between the base station and user $u$, and $\bm{n}_u$ is the AWGN vector, modeled as a zero-mean circularly symmetric complex Gaussian process with covariance matrix $N_0 \bm{I}$. The channel is assumed to follow block Rayleigh fading, with $h_u \sim \mathcal{CN}(0, \sigma_{h_u}^2)$, independently across users.

In accordance with NOMA principles, users are sorted by ascending channel gain: $|h_1|^2 < |h_2|^2 < \cdots < |h_U|^2$, and power is allocated inversely, i.e., $P_1 > P_2 > \cdots > P_U$, so that users with weaker channels are assigned higher power \cite{NOMA}. In this work, we adopt a fixed power allocation strategy \cite{FPA-NOMA}, where the power for each user is recursively defined using a power allocation factor $0 < \beta < 1$:
\begin{align}
P_t &= \sum_{u=1}^U P_u, \quad P_0 = P_t, \nonumber \\
P_u &= \beta P_{u-1}, \quad \forall u \in [1, U],
\label{eq-FPA-NOMA}
\end{align}
where $P_t$ denotes the total transmit power of the base station. A smaller $\beta$ results in a larger power difference among users. Although adaptive power allocation strategies could further optimize performance, such techniques are beyond the scope of this paper. Meanwhile, adopting a recursive power allocation strategy enables a systematic analysis of how power disparities among users influence the PAPR performance of the proposed scheme. This approach facilitates clear observation of the power impact by adjusting allocation factors across users. Since PAPR is a key metric for assessing the feasibility of waveform designs in practical NOMA deployments, this strategy is particularly valuable.

At the receiver side, the strategy of successive interference cancellation (SIC) is employed. User $u$ attempts to decode and cancel the signals of all lower-priority users $1, 2, \cdots, u-1$ before decoding its own signal. After canceling users $1$ through $q-1$, the residual signal available at user $u$ for decoding the $q^\text{th}$ signal is modeled as:
\begin{align}
\bm{r}_{u,q} &= \underbrace{h_u \sqrt{P_q} \bm{s}_q}_{\text{the desried symbol}} + \underbrace{\sum_{n=q+1}^U h_u \sqrt{P_n} \bm{s}_n}_{\text{the remaining IUI}}     \nonumber \\
                     & \quad {}+  \underbrace{\sum_{n=1}^{q-1} h_u \sqrt{P_n} \Delta_\text{SIC}}_{\text{the imperfect SIC}}  + \, \bm{n}_u, 
\label{eq-NOMA-SIC}
\end{align}
where $\Delta_\text{SIC}$ denotes the residual interference due to imperfect cancellation of the previously decoded signals.

In conventional systems, the maximum likelihood sequence detection based on the Viterbi algorithm is used to extract the desired modulated symbols from $\bm{r}_{u,q}$. The original data is then recovered via despreading. However, this separate processing introduces performance degradation.

In contrast, the proposed IM-CPM-SS and CIM-CPM-SS schemes utilize a joint demodulation and despreading mechanism, wherein the received signal is processed in a unified step to extract both the index and modulated bits. This joint operation leverages the quasi-orthogonality of the codewords in $\bm{Z}$, enabling more accurate and efficient bit recovery compared to conventional schemes that perform demodulation and despreading separately. Furthermore, in NOMA systems where users are distinguished based on power levels, assigning distinct quasi-orthogonal sequences to different users significantly mitigates inter-user interference and improves the effectiveness of SIC. As such, the joint demodulation and despreading approach offers substantial advantages in multi-user NOMA scenarios by enhancing interference suppression and decoding efficiency.

\section{Performance analysis and Impact of Power Amplifier Nonlinear Distortion}\label{Section-IV}
This section presents the analytical performance of the proposed IM-CPM-SS and CIM-CPM-SS modulation schemes. We begin by deriving the BER under AWGN and Rayleigh fading channels. Next, we evaluate the spectral efficiency, energy efficiency, and computational complexity of the proposed methods. Finally, we analyze the PAPR and assess the impact of power amplifier nonlinear distortion.

\subsection{BER Analysis}
\label{Section-IV-A}
\subsubsection{\underline{IM-CPM-SS Scheme}}
For IM-CPM-SS, the BER under noncoherent detection of orthogonal signals over an AWGN channel is given by \cite[Eq.~(4.5-44)]{Digital-Communication}:
\begin{equation}\label{eq-scheme1-ber-a}
       P_b = \frac{2^{b_c-1}}{2^{b_c}-1} \sum_{k=1}^{2^{b_c}-1} \frac{(-1)^{k+1}}{k+1} \binom{2^{b_c}-1}{k} e^{-\frac{k}{k+1} b_c\frac{E_b}{N_0}}, 
\end{equation}
where $E_b / N_0$ denotes the signal-to-noise ratio (SNR) per bit, and $2^{b_c}$ counts the number of orthogonal signals. Over a frequency-flat Rayleigh fading channel, the BER becomes \cite[Eq.~(13.4-50)]{Digital-Communication}:
\begin{equation}\label{eq-scheme1-ber-r}
      P_b = \frac{2^{b_c-1}}{2^{b_c}-1} \sum_{k=1}^{2^{b_c}-1} \frac{(-1)^{k+1}\binom{2^{b_c}-1}{k}}{1+k+k \overline{\gamma}_b}, 
\end{equation}
where $\overline{\gamma}_b = \mathbb{E}(|h|^2) E_b / N_0$ denotes the average SNR, with $h \sim \mathcal{CN}(0,1)$ being the channel coefficient.

\subsubsection{\underline{CIM-CPM-SS Scheme}}
For CIM-CPM-SS, each symbol includes both index bits (mapped to a spreading codeword) and modulated bits (mapped to a PSK symbol). Let $P_d$ denote the probability of incorrect codeword detection. Then, the BER for the index bits is \cite[Eq.~(4.5-47)]{Digital-Communication} 
\begin{align}
   P_{c} = \frac{2^{b_c-1}}{2^{b_c}-1} P_{d},
\label{eq-CIM-CPM-SS-pc}
\end{align}
and the overall BER is a weighted average:
\begin{align}
   P_{T} = \frac{b_c}{b_c+b_m} P_{c} +  \frac{b_m}{b_c+b_m} P_{m},
\label{eq-CIM-CPM-SS-pT}
\end{align}
where $P_m$ denotes the BER of the modulated bits. Errors in the modulated bits arise from two distinct cases:
\begin{itemize}
\item When the spreading codeword is correctly identified, but the corresponding PSK symbol is erroneously demodulated;
\item When the spreading codeword is incorrectly detected, in which case the demodulated PSK symbol has a $50\%$ probability of being correct, assuming Gray-coded constellation mapping.
\end{itemize}
Thus, $P_m$ can be approximated as:
\begin{align}
   P_{m} = P_\text{PSK}(1-P_{c}) +  \frac{1}{2} P_{c}.
\label{eq-CIM-CPM-SS-pm}
\end{align}

The codeword detection error $P_d$ can be computed as:
\begin{align}
   P_{d} &= 1 - \mathrm{Pr} \left( \max_{n,\ n\neq \hat{n}} \{|D_{i,n}|\} < |D_{i,\hat{n}}| \Big|h_i\right) \nonumber \\
              &= 1 - \mathrm{Pr} \left( |D_{i,1}| < |D_{i,\hat{n}}|, \cdots , |D_{i,K-1}| < |D_{i,\hat{n}}|\Big|h_i\right)   \nonumber  \\
              &= 1 - \int_{0}^{\infty} \left[\mathrm{Pr}(\vert D_{i,n} \vert < x | h_i)\right]^{2^{b_c}-1} f_{\vert D_{i,\hat{n}} \vert}(x) \, {\rm d}x,
\label{eq-scheme2-ped-v}
\end{align}
where $|D_{i,\hat{n}}|$ follows a Rician distribution with mean $u_1 = E_s |h_i|^2$ and variance $\sigma_1^2 = E_s N_0 |h_i|^2 / 2$, while $|D_{i,n}|$ for $n \ne \hat{n}$ follows a Rayleigh distribution with variance $\sigma_2^2 = E_s N_0 |h_i|^2 / 2$. Their respective probability density function (PDF) and cumulative distribution functions (CDF) are:
\begin{align}
f_{|D_{i,\hat{n}}|}(x) &= \frac{x}{\sigma_1^2} \exp\left(-\frac{x^2 + u_1^2}{2\sigma_1^2}\right) I_0\left(\frac{x u_1}{\sigma_1^2} \right), \, x \geq 0, \label{eq-RicianPDF} \\
F_{|D_{i,n}|}(x) &= 1 - \exp\left(-\frac{x^2}{2\sigma_2^2}\right), \, x \geq 0. \label{eq-RayleighCDF}
\end{align}

Let the instantaneous SNR be defined as $\gamma_b = |h_i|^2 E_b / N_0$, with an exponential PDF:
\begin{align}
  f_{\gamma_b}(\gamma_b) = \frac{1}{\overline{\gamma}_b} \exp\left(-\frac{\gamma_b}{\overline{\gamma}_b}\right), 
\label{eq-scheme2-rb}
\end{align}
where $\overline{\gamma}_b = \mathbb{E}(|h_i|^2) E_b / N_0$ is the average SNR. Substituting \eqref{eq-RicianPDF}-\eqref{eq-scheme2-rb} into \eqref{eq-scheme2-ped-v}, we obtain
\begin{align}
  P_{d} &= \int_{0}^{\infty} \left\{ 1-\int_{0}^{\infty}  \left(1-\exp\left(-\frac{x^2}{b_m \gamma_b N_0^2}\right)\right)^{2^{b_c}-1} \hspace{-0.5em} \frac{2x}{b_m \gamma_b N_0^2}  \right.   \nonumber  \\
             & \quad \left.  \times \exp\left(-\frac{(x^2 + b_m^2 \gamma_b^2 N_0^2)}{b_m \gamma_b N_0^2}\right) I_0\left( \frac{2x}{N_0} \right) \, {\rm d}x\right\}f_{\gamma_b}(\gamma_b) \, {\rm d}\gamma_b. 
\label{eq-scheme2-ped-final}
\end{align}

For $M$-ary PSK modulation, its BER depends on $b_m = \log_2 M$. If the modulation order $M \leq 4$, the BER is given by \cite[Eq.~(4.3-13)]{Digital-Communication} 
\begin{equation}
  P_\text{PSK} = \int_{0}^{\infty} \frac{1}{2} \text{erfc}(\sqrt{\gamma_b}) f_{\gamma_b}(\gamma_b) \, {\rm d}\gamma_b.
\label{eq-PSK-QPSK}
\end{equation}
If $M > 4$, the BER can be approximated as \cite[Eq.~(4.3-19)]{Digital-Communication}
\begin{align}
  P_\text{PSK} = \int_{0}^{\infty} \hspace{-0.25em} \frac{1}{b_m} \text{erfc}\left(\sqrt{b_m \gamma_b \sin^2 \left( \frac{\pi}{M} \right) } \right) f_{\gamma_b}(\gamma_b) \, {\rm d}\gamma_b.  
\label{eq-PSK-MPSK}
\end{align}

Finally, the total BER $P_T$ of CIM-CPM-SS is obtained by substituting \eqref{eq-CIM-CPM-SS-pc}, \eqref{eq-CIM-CPM-SS-pm}, and \eqref{eq-PSK-QPSK}–\eqref{eq-PSK-MPSK} into \eqref{eq-CIM-CPM-SS-pT}.

\subsection{Spectral Efficiency}
\label{Section-IV-B}
The spectral efficiency, defined as the number of transmitted bits per unit time per unit bandwidth, is given by
\begin{align} \label{eq-SE}
  \eta_{\text{SE}} = \frac{R_b}{B} = \frac{b_t / T_s}{(2^{\text{SF}} - 1) / T_s} = \frac{b_t}{2^{\text{SF}} - 1},
\end{align}
where $T_s$ denotes the symbol duration, $b_t$ is the number of transmitted bits per symbol, $R_b = b_t / T_s$ is the bit rate, and $B = (2^{\text{SF}} - 1) / T_s$ represents the occupied bandwidth corresponding to a spreading factor of $\text{SF}$.

The following schemes are compared in this paper:
\begin{enumerate}
   \item DSSS-CPM-sep: Conventional CPM-SS with separate demodulation and despreading.
   \item DSSS-CPM: CPM-SS with joint demodulation and despreading \cite{Iot-zhangruiqi}.
   \item IM-CPM-SS and CIM-CPM-SS: The proposed schemes with joint detection.
   \item IM-CPM-SS-sep: IM-CPM-SS with separate demodulation and despreading.
   \item CIM: Conventional CIM scheme with identical spreading codes for in-phase and quadrature branches \cite{CIM}.
\end{enumerate}

\begin{table}[!t]
		\renewcommand\arraystretch{1.5}
		\begin{center}
			\caption{Spectral Efficiency Comparisons}
			\label{tab-data-rate}
			\begin{tabular}{!{\vrule width1.2pt} c | c !{\vrule width1.2pt}}
				\Xhline{1.2pt} 
				{\bf Tx Scheme} & {\bf Spectral Efficiency (bps/Hz)} \\
				\Xhline{1.2pt} 
				DSSS-CPM-sep, DSSS-CPM &  $1/(2^{\text{SF}} - 1)$ \\
        \hline
				IM-CPM-SS-sep, IM-CPM-SS & $b_c/(2^{\text{SF}} - 1)$ \\
        \hline
				CIM, CIM-CPM-SS & $(b_c + b_m)/(2^{\text{SF}} - 1)$ \\
        \Xhline{1.2pt} 
			\end{tabular}
		\end{center}
\end{table}

Table~\ref{tab-data-rate} summarizes the spectral efficiency of the aforementioned transmission schemes, incorporating the effects of different spreading factors. The evaluation methodology aligns with prior studies, such as \cite{LoRa-2019, DM-Mode-TDM-CSS}. Assuming equal symbol intervals and setting $b_c = b_m = 2$, the DSSS-CPM-sep and DSSS-CPM schemes each achieve a spectral efficiency of $1$ bit per symbol per unit bandwidth, owing to their conventional spreading and modulation structures. In contrast, the IM-CPM-SS-sep and IM-CPM-SS schemes improve efficiency by a factor of two, transmitting $2$ bits per symbol per unit bandwidth through index-based codeword selection. The CIM and CIM-CPM-SS schemes further boost the spectral efficiency to $4$ bits per symbol per unit bandwidth by jointly encoding both index and modulation bits.

\subsection{Energy Efficiency}
\label{Section-IV-C}
In the CIM-CPM-SS scheme, only $b_m$ out of the total $b_g = b_c + b_m$ transmitted bits are directly modulated using $2^{b_m}$-ary PSK. The remaining $b_c$ bits are implicitly transmitted through the index selection of spreading codewords. As a result, for the same number of transmitted bits $b_g$, the percentage of energy savings achieved by CIM-CPM-SS over the conventional DSSS-CPM scheme \cite{Iot-zhangruiqi} is given by:
\begin{align} \label{eq-Ee}
	 \eta_{_\text{EE}} =  \frac{b_c}{b_c+b_m} \times 100\%.  
\end{align}

\subsection{Computational Complexity Analysis}
\label{Section-IV-D}
In this subsection, we assess the computational complexity of the proposed schemes, using the number of complex multiplications as the primary metric.
\begin{enumerate}
	\item \underline{IM-CPM-SS-sep:} At the receiver, this scheme performs CPM demodulation followed by despreading. The number of CPM phase states is $2h_d M^{L-1}$, and each state has $M$ transition branches. Each branch computation requires $P$ complex multiplications. Therefore, the complexity of maximum likelihood sequence detection based on the Viterbi algorithm is $\mathcal{O}(2h_d M^{L} P N N_s)$, where $N$ is the spreading code length and $N_s$ is the number of CPM-SS symbols. The subsequent despreading process requires correlation with $K = 2^{b_c}$ candidate codewords, resulting in an additional complexity of $\mathcal{O}(K N N_s / 4)$. Hence, the total complexity is $\mathcal{O}(2h_d M^{L}NPN_s + KNN_s/4)$. 
	\item \underline{IM-CPM-SS:} This scheme avoids maximum likelihood sequence detection based on the Viterbi algorithm and performs only correlation with $K$ modulated codewords of length $NP$, resulting in a total complexity of $\mathcal{O}(K N P N_s)$.
	\item \underline{CIM-CPM-SS:} The despreading step is the same as in IM-CPM-SS, with complexity $\mathcal{O}(K N P N_s)$. An additional $2^{b_m}$-ary PSK symbol detection step contributes a complexity of $\mathcal{O}(M_{\text{PSK}} N_s)$, where $M_{\text{PSK}} = 2^{b_m}$. Thus, the total complexity is $\mathcal{O}(K N P N_s + M_{\text{PSK}} N_s)$. 
\end{enumerate}
To provide more concrete insights, we consider the typical system parameters used throughout this paper: binary CPM with $M = 2$, $h_d = 2$, memory length $L = 1$, and spreading code length $N = 2^{\text{SF}} - 1 \approx 2^{\text{SF}}$. Under these assumptions, the complexities simplify as follows:
\begin{itemize}
	\item IM-CPM-SS-sep: $\mathcal{O}(2^{\text{SF}+3}PN_s + 2^{b_c+\text{SF}-2}N_s)$
	\item IM-CPM-SS: $\mathcal{O}(2^{b_c+\text{SF}}PN_s)$
	\item CIM-CPM-SS: $\mathcal{O}(2^{b_c+\text{SF}}PN_s + 2^{b_m} N_s)$
\end{itemize}
These results reveal that, for small values of $b_c$, the IM-CPM-SS-sep scheme entails higher complexity than IM-CPM-SS for the same number of transmitted bits per symbol. As $b_c$ increases, the complexity of IM-CPM-SS eventually surpasses that of the IM-CPM-SS-sep scheme. Therefore, by flexibly adjusting the transmission parameters $(b_c, b_m)$, the CIM-CPM-SS scheme can match the complexity of IM-CPM-SS while offering more flexible trade-offs in terms of spectral efficiency and error performance.

\subsection{Impact of Power Amplifier Nonlinear Distortion}
\label{Section-IV-E}
To ensure sufficient received signal power, the power amplifier at the transmitter is typically operated near its saturation region to maximize efficiency \cite{HPA}. However, operating in this regime introduces nonlinear distortion, primarily due to the amplitude nonlinearity of the power amplifier.

Let the input signal be represented as $\bm{s} = \bm{s}_a \exp(j \bm{\phi}_p)$, where $\bm{s}_a$ and $\bm{\phi}_p$ denote the signal's amplitude envelope and phase, respectively. The output of a nonlinear power amplifier can be modeled as
\begin{align}
  \bm{s}_\text{PA} = F_a(\bm{s}_a)e^{j F_p(\bm{s}_a)} e^{j \bm{\phi}_p}, 
\label{eq-PA}
\end{align}
where the functions $F_a(\cdot)$ and $F_p(\cdot)$ reflect the nonlinear amplitude modulation-amplitude modulation (AM-AM) and amplitude modulation-phase modulation (AM-PM) characteristics, respectively. In this work, we adopt the widely used Rapp model \cite{CPM-NOMA, Rapp} to characterize the AM-AM nonlinearity while ignoring AM-PM distortion, i.e., $F_p(\bm{s}_a) = 0$.

According to the Rapp model, the amplitude transfer function is given by
\begin{align}
  F_a(\bm{s}_a) = \frac{G_0 \bm{s}_a}{\left( 1 + \left( \frac{G_0 \bm{s}_a}{S_\text{sat}}\right)^{2p} \right)^{\frac{1}{2p}}},
\label{eq-Rapp}
\end{align}
where $G_0$ is the small-signal gain, $S_\text{sat}$ is the saturation voltage (normalized to $1$ in this study), and $p$ is the smoothness parameter that governs the transition from the linear region to saturation (set to $p=2$).

To mitigate nonlinear effects, IBO is often used to ensure that the power amplifier operates within its quasi-linear region. However, this comes at the cost of reduced power efficiency. A critical metric related to IBO is the PAPR, defined as
\begin{align}
  \text{PAPR} = 10 \log_{10} \left( \frac{\max \left(\vert \bm{s} \vert^2 \right)}{\mathbb{E}\left( \vert \bm{s} \vert^2\right)}\right) \, \si{dB}, 
\label{eq-PAPR}
\end{align}
where the numerator represents the instantaneous peak power, and the denominator denotes the average power of the transmitted signal. A lower PAPR enables operation with reduced IBO, thus improving the power efficiency of the power amplifier.

To statistically characterize the PAPR behavior, the complementary cumulative distribution function (CCDF) is widely used. The CCDF describes the probability that the PAPR exceeds a certain threshold and serves as an effective metric for comparing the PAPR performance across different modulation schemes. Simulation results illustrating these comparisons are presented in the next section.

\begin{figure*}[!t]
    \centering
    \subfigure[IM-CPM-SS]{
		\includegraphics[width=0.3\linewidth]{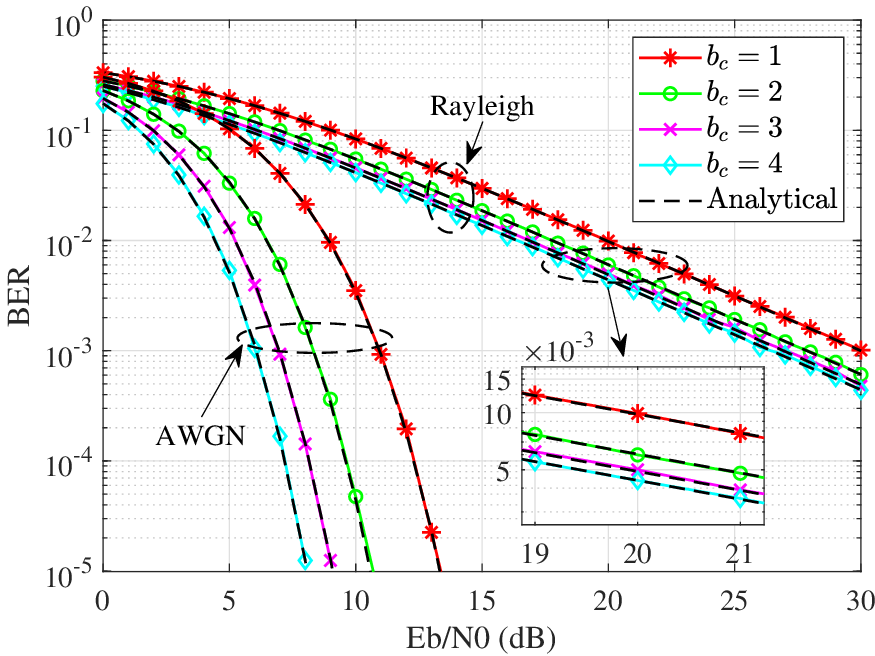} \label{subfig-im-ber-proposed-schemes}}
	\subfigure[CIM-CPM-SS]{
		\includegraphics[width=0.3\linewidth]{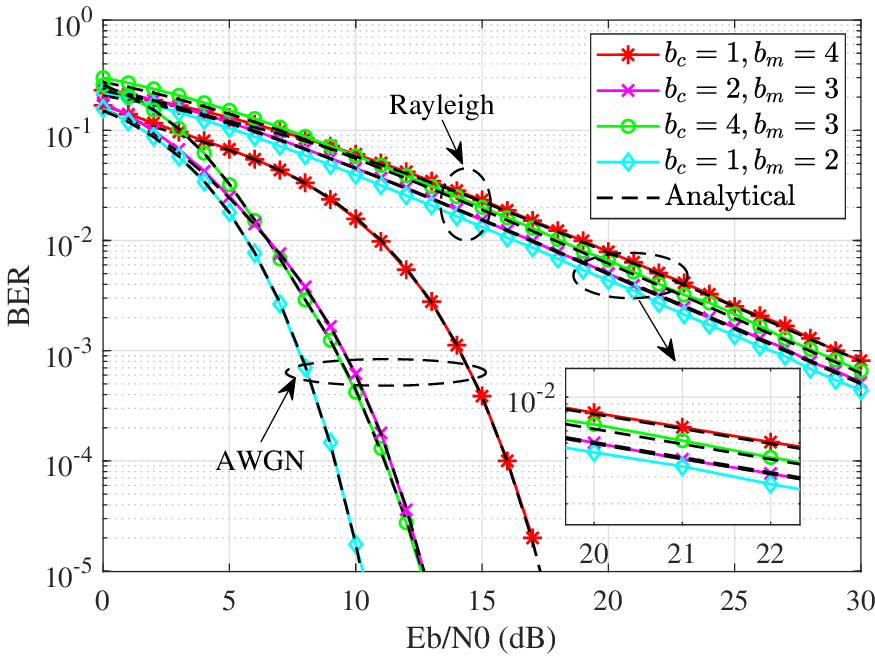} \label{subfig-cim-ber-proposed-schemes}}
	\subfigure[Different shaping pulses]{
		\includegraphics[width=0.3\linewidth]{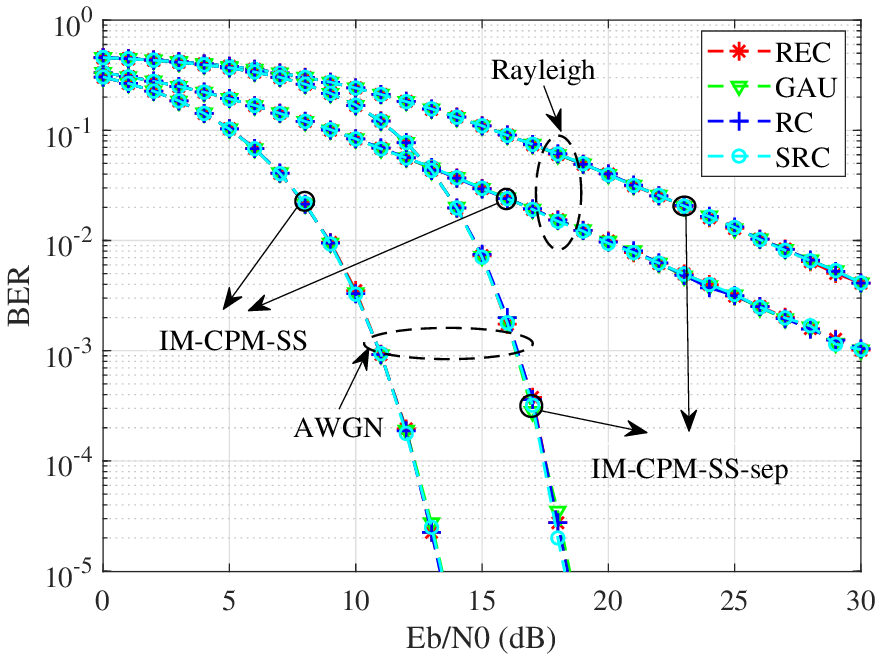} \label{subfig-pulse-ber-proposed-schemes}}
	\caption{BER vs. $E_b/N_0$ of two proposed schemes for different parameters over AWGN and Rayleigh channels (with a fixed spreading factor $\text{SF}=6$).}
\label{fig-ber-proposed-schemes}
\end{figure*}

\begin{figure*}[!t]
    \centering
    \subfigure[Proposed schemes]{
		\includegraphics[width=0.3\linewidth]{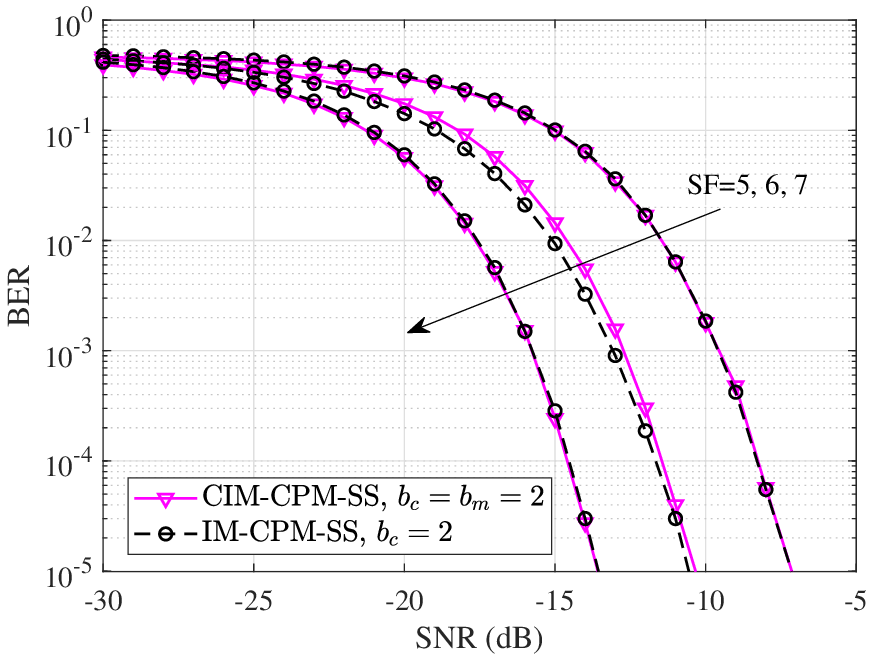} \label{subfig-diff-SF}}
	\subfigure[Different detection schemes]{
		\includegraphics[width=0.3\linewidth]{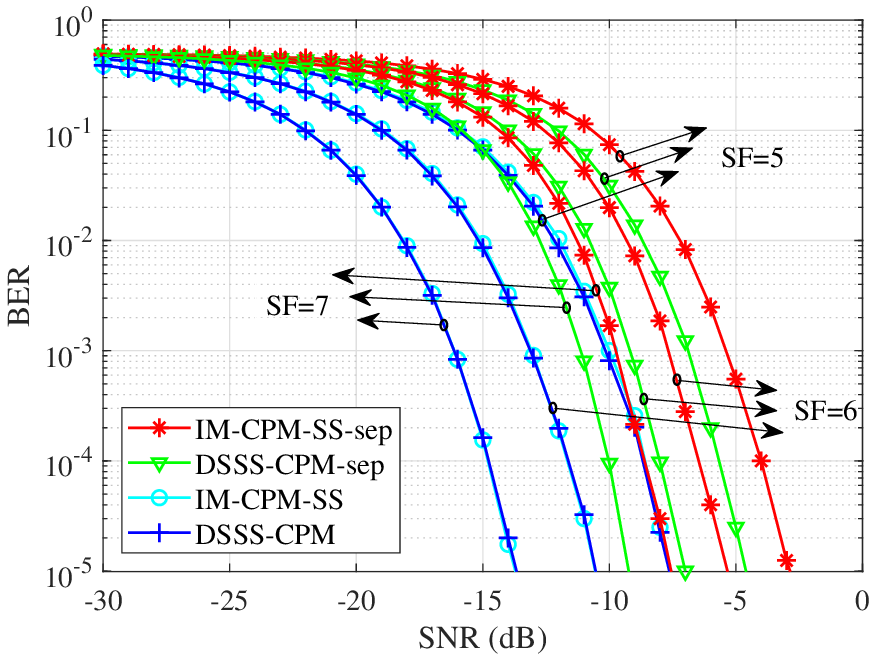} \label{subfig-diff-detect}}
	\subfigure[Different schemes ($\text{SF}=6$)]{
		\includegraphics[width=0.3\linewidth]{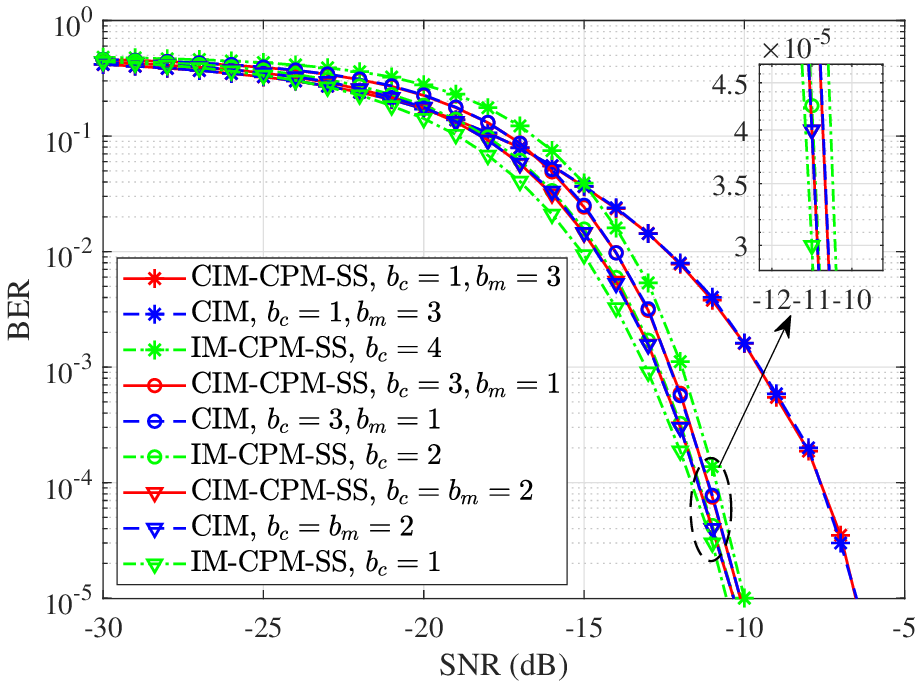} \label{subfig-diff-scheme}}
	\caption{BER vs. SNR of the different schemes over AWGN channels (with varying spreading factors).}
\label{fig-ber-diff-schemes-snr}
\end{figure*}

\section{Simulation Results and Discussions} 
\label{Section-V}
This section presents and discusses simulation and analytical results for the proposed IM-CPM-SS and CIM-CPM-SS schemes over AWGN and Rayleigh fading channels. For comparison, we include several benchmark schemes: DSSS-CPM-sep, DSSS-CPM, IM-CPM-SS-sep, and CIM. All simulation experiments are conducted using MATLAB R2024a.

The pulse shaping filters used in the experiments are characterized as follows: the GAU pulse is specified by its time-bandwidth product $BT$, while the SRC and square-root raised cosine (SRRC) pulses are defined by their roll-off factor $\alpha$. The symmetric SRRC pulse used in CIM-based modulation is further truncated by a length parameter $L_t$ to ensure finite duration. A summary of the key simulation parameters is provided in Table~\ref{tab-sim-params} for clarity and reproducibility.

\subsection{BER of Point-to-Point Transceiver}
 \label{Section-V-A}
Fig.~\ref{subfig-im-ber-proposed-schemes} shows the simulated BER performance of the IM-CPM-SS scheme versus $E_b/N_0$ in \si{dB}, while Fig.~\ref{subfig-cim-ber-proposed-schemes} presents that of the CIM-CPM-SS scheme over AWGN and Rayleigh fading channels with a fixed spreading factor $\text{SF} = 6$. As observed, the BER performance of the proposed scheme remains robust even when a small spreading factor, such as $\text{SF}=6$, is used. Although the orthogonality among sequences in the codebook $\bm{Z}$ is limited in this case (e.g., the $m$-sequence length is $N = 2^{\text{SF}} - 1 = 63$ and the codeword length is $NP = 252$ with $P=4$ samples per chip), the detection process still benefits from a high contrast in cross-correlation magnitudes. Specifically, the normalized cross-correlation magnitude between the received signal and the correct codeword is close to~$1$, while the corresponding value for any incorrect codeword is on the order of $1/(NP) = 1/252$. This significant disparity ensures accurate detection of the transmitted codeword, thereby yielding favorable BER performance even when the codewords are not fully orthogonal. This observation highlights the robustness of the proposed scheme in low-SF scenarios. Additionally, in both subfigures, the simulated BER closely matches the analytical expressions, confirming the accuracy of our theoretical derivations.

For the IM-CPM-SS scheme, Fig.~\ref{subfig-im-ber-proposed-schemes} illustrates that the BER improves as the number of index bits $b_c$ increases. This is attributed to the larger codeword search space associated with a higher $b_c$, which improves detection performance, as indicated by \eqref{eq-scheme1-ber-a} and \eqref{eq-scheme1-ber-r}. However, the performance gain diminishes with larger $b_c$ due to diminishing returns in Euclidean distance diversity.

In contrast, as seen in Fig.~\ref{subfig-cim-ber-proposed-schemes}, the BER of CIM-CPM-SS deteriorates with increasing $b_m$ (modulated bits), while keeping $b_c$ fixed. This is because increasing $b_m$ reduces the Euclidean distance between PSK symbols, making them more susceptible to noise. Although the BER also slightly degrades with increasing $b_c$, increasing $b_c$ is generally more beneficial, as it improves spectral efficiency while preserving more robust symbol separation in the modulation domain.

\begin{table}[!t]
    \renewcommand\arraystretch{1.3}
    \centering
    \caption{Parameter Setting in Simulation Experiments}
    \label{tab-sim-params}
    \begin{tabular}{!{\vrule width1.2pt} >{\centering\arraybackslash}m{0.38\linewidth}|>{\centering\arraybackslash}m{0.4\linewidth} !{\vrule width1.2pt}}
       \Xhline{1.2pt} 
        \textbf{Parameter} & \textbf{Value} \\
       \Xhline{1.2pt} 
        CPM modulation index ($h$) & $1/2$ \\
        \hline
        CPM memory length ($L$) & $1$ \\
        \hline
        CPM modulation order ($M$) & $2$ \\
        \hline
        Oversampling factor ($P$) & $4$ \\
        \hline
        GAU pulse & $BT=0.3$ \\
        \hline
        SRC pulse & $\alpha=0.3$ \\
        \hline
        SRRC pulse & $L_t=6T$ \\
        \hline
        NOMA channel variance & $\sigma_{h_1}^2 = 1, \, \sigma_{h_2}^2 = 2, \, \sigma_{h_3}^2 = 3$ \\
        \hline
        NOMA transmit power & $P_t = 1 $ \\
       \Xhline{1.2pt} 
    \end{tabular}
\end{table}

Fig.~\ref{subfig-pulse-ber-proposed-schemes} examines the BER of the IM-CPM-SS and IM-CPM-SS-sep schemes under various pulse shaping filters (e.g., REC, GAU, SRC, SSRC) with $b_c = 1$, under both AWGN and Rayleigh fading channels. The BER remains consistent across different pulse shapes, despite differences in their power spectral densities. Hence, the REC pulse is selected for the remainder of the simulations, aligning with the IEEE 802.15.4 standard for minimum shift keying physical layer implementations \cite{2024-IEEE-standard}.

Fig.~\ref{fig-ber-diff-schemes-snr} plots the BER versus SNR, in contrast to Fig.~\ref{fig-ber-proposed-schemes}, where $E_b/N_0$ is the horizontal axis. Since the energy per CPM-SS symbol is fixed, transmitting more bits per symbol (i.e., higher $b_g$) leads to lower energy per bit, resulting in increased BER at the same SNR. This clarifies the distinction between evaluating BER using $E_b/N_0$ and SNR.

Fig.~\ref{subfig-diff-SF} compares the BER of the CIM-CPM-SS scheme (with $b_c = b_m = 2$) and the IM-CPM-SS scheme (with $b_c = 2$) under different spreading factors ($\text{SF} = 5, 6, 7$). Each additional unit of $\text{SF}$ yields an approximate 3 \si{dB} SNR gain at the same BER, highlighting the sensitivity improvement achieved by increasing the spreading factor. Both schemes exhibit nearly identical BERs under the same $\text{SF}$, but CIM-CPM-SS offers a higher spectral efficiency by incorporating modulated bits.

Fig.~\ref{subfig-diff-detect} compares the BERs of the IM-CPM-SS, DSSS-CPM-sep, and DSSS-CPM schemes under various spreading factors. With fixed $b_c = 1$, both IM-CPM-SS and DSSS-CPM exhibit similar performance and significantly outperform their respective “-sep” variants. Specifically, the proposed IM-CPM-SS achieves SNR gains of approximately $3$, $3.5$, and $4.5$ \si{dB} over DSSS-CPM-sep, and $4.5$, $5$, and $6$ \si{dB} over IM-CPM-SS-sep, at $\text{BER} = 10^{-4}$ for $\text{SF} = 5, 6, 7$, respectively. Notably, for the schemes with separate demodulation and despreading, each increment in $\text{SF}$ yields only a 2 \si{dB} SNR gain, less than the 3~\si{dB} observed for the joint schemes, demonstrating the inefficiency of separation.

Fig.~\ref{subfig-diff-scheme} compares the BERs of the CIM-CPM-SS, CIM, and IM-CPM-SS schemes for a fixed $\text{SF} = 6$. The CIM-CPM-SS and CIM schemes show nearly identical BERs due to their similar structure. For the IM-CPM-SS scheme, increasing $b_c$ by $1$ results in a 0.5 \si{dB} degradation at $\text{BER} = 10^{-4}$. Among configurations transmitting 4 bits per symbol, the best performance is observed for the setting with $b_c = b_m = 2$, due to the optimal balance between index coding and PSK symbol separation. The configuration with $b_c = 3$ and $b_m = 1$ outperforms that with $b_c = 1$ and $b_m = 3$, again underscoring the benefits of maintaining a strong Euclidean distance between modulation symbols. Overall, the CIM-CPM-SS scheme can flexibly trade off between BER and throughput, outperforming IM-CPM-SS under matched bit rates.

\subsection{Tradeoff Between Spectral Efficiency and Complexity}
\begin{figure}[!t]
   \centering
   \subfigure[Spectral efficiency]{
		\includegraphics[width=0.45\hsize, height=0.65\hsize]{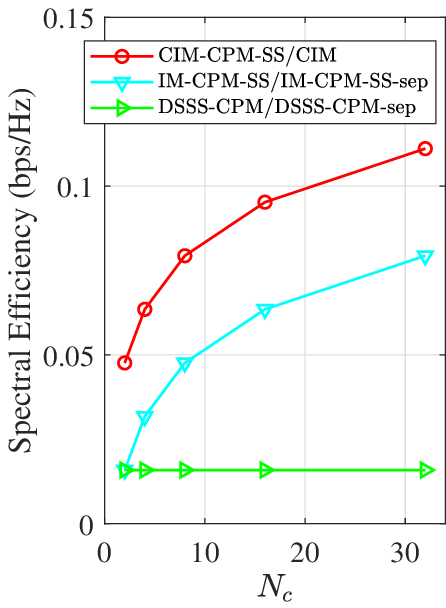} \label{subfig-comparison-SE}}
	\subfigure[Computational complexity]{
		\includegraphics[width=0.45\hsize, height=0.65\hsize]{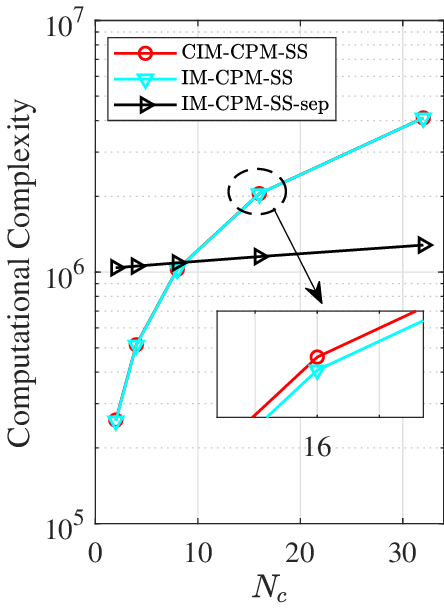} \label{subfig-comparison-complexity}}
   \caption{Spectral efficiency and computational complexity comparisons of the proposed and benchmark schemes with varying number of usable codewords ($N_c$).}
   \label{fig-tradeoff-SE-complexity}
\end{figure}

Fig.~\ref{subfig-comparison-SE} illustrates the spectral efficiency of the proposed IM-CPM-SS and CIM-CPM-SS schemes, in comparison with several benchmarks: DSSS-CPM-sep, DSSS-CPM, CIM, and IM-CPM-SS-sep. For the parameters $M=4$ and $\text{SF}=6$, CIM-CPM-SS and CIM achieve the highest spectral efficiency, followed by IM-CPM-SS and IM-CPM-SS-sep. The DSSS-based schemes exhibit the lowest efficiency due to the absence of index modulation. Notably, for both IM-CPM-SS and CIM-CPM-SS, the spectral efficiency increases linearly with the number of index bits $b_c = \log_2(N_c)$, whereas the efficiency of DSSS-based schemes remains constant.

Fig.~\ref{subfig-comparison-complexity} compares the computational complexity under $\text{SF}=6$, $P=4$, and $N_s = 500$. CIM-CPM-SS introduces slightly higher complexity than IM-CPM-SS owing to the additional data bits conveyed through PSK modulation. When $b_c$ is small, IM-CPM-SS-sep shows higher complexity due to separate detection steps. However, as $b_c$ increases, the complexity of IM-CPM-SS and CIM-CPM-SS surpasses that of the separated scheme because of the larger joint search space.

In summary, IM-CPM-SS and CIM-CPM-SS provide a favorable tradeoff between efficiency and complexity. IM-CPM-SS is preferable when $b_c$ is small, offering moderate efficiency gains with lower complexity. CIM-CPM-SS delivers higher spectral efficiency at the cost of modest complexity overhead. In contrast, DSSS-based schemes are computationally simpler but exhibit limited scalability in spectral efficiency.

\subsection{PAPR of NOMA Downlink Transmission}
 \label{Section-V-B}
 
\begin{figure}[!t]
   \centering
   \subfigure[Two-user case]{
		\includegraphics[width=0.45\hsize, height=0.65\hsize]{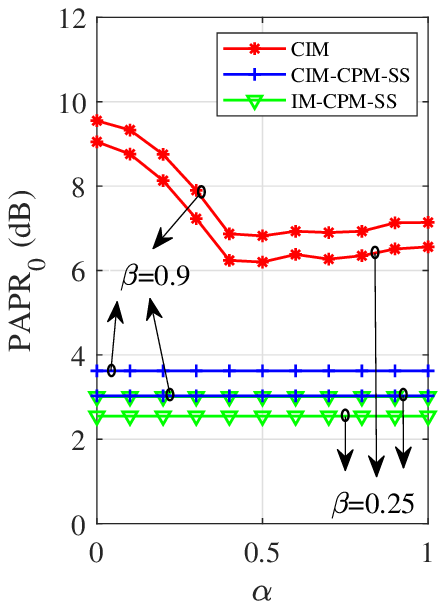} \label{subfig-papr-noma-K2}}
	\subfigure[Three-user case]{
		\includegraphics[width=0.45\hsize, height=0.65\hsize]{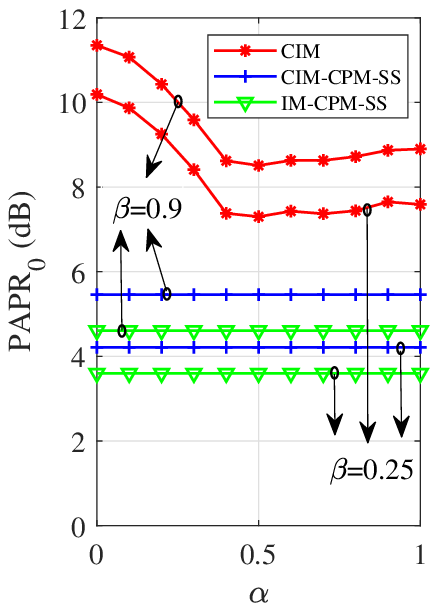} \label{subfig-papr-noma-K3}}
   \caption{PAPR vs. roll-off factor ($\alpha$) for IM-CPM-SS, CIM-CPM-SS and CIM with different power allocation factors ($\beta$) under NOMA scenario.}
   \label{fig-papr-noma}
\end{figure}

Fig.~\ref{fig-papr-noma} compares the PAPR performance of the IM-CPM-SS ($b_c=4$), CIM-CPM-SS ($b_c = b_m = 2$), and CIM ($b_c = b_m = 2$) schemes versus the roll-off factor $\alpha$, for both two-user and three-user NOMA scenarios under different power allocation settings. The vertical axis, $\text{PAPR}_0$ in the unit of dB, denotes the PAPR value computed by \eqref{eq-PAPR} corresponding to the percentile where the CCDF drops below $10^{-4}$.

For the CIM scheme, $\text{PAPR}_0$ decreases as the roll-off factor $\alpha$ increases, eventually stabilizing once $\alpha$ exceeds $0.5$. This trend suggests that the envelope fluctuations of the SRRC pulse used in CIM modulation are more severe at smaller roll-off factors and are mitigated as $\alpha$ increases. In the two-user case, the minimum $\text{PAPR}_0$ values are $6.20$ \si{dB} for $\beta = 0.25$ (with $P_1 = 0.8$ and $P_2 = 0.2$), and $6.82$ \si{dB} for $\beta = 0.9$ (with $P_1 = 0.5263$ and $P_2 = 0.4737$). For the three-user case, the minimum values increase to $7.30$ \si{dB} for $\beta = 0.25$ (with $P_1 = 0.7619$, $P_2 = 0.1905$, and $P_3 = 0.0476$), and $8.51$ \si{dB} for $\beta = 0.9$ (with $P_1 = 0.3690$, $P_2 = 0.3321$, and $P_3 = 0.2989$). These reference values are adopted in the subsequent simulation experiments for consistent performance evaluation of the CIM-CPM-SS and IM-CPM-SS schemes.

In contrast, the $\text{PAPR}_0$ values of IM-CPM-SS and CIM-CPM-SS remain nearly constant with respect to the roll-off factor, and are significantly lower than those of CIM. Specifically, under the two-user setting with $\beta = 0.25$, $\text{PAPR}_0$ values are $2.55$ \si{dB} for IM-CPM-SS and $3.03$ \si{dB} for CIM-CPM-SS. When the power allocation becomes more balanced ($\beta = 0.9$), these values increase to $3.01$ \si{dB} and $3.62$ \si{dB}, respectively. For the three-user scenario, the $\text{PAPR}_0$ values are $3.60$ \si{dB} (IM-CPM-SS) and $4.21$ \si{dB} (CIM-CPM-SS) for $\beta = 0.25$, increasing to $4.62$ \si{dB} and $5.46$ \si{dB} for $\beta = 0.9$. These results demonstrate that a larger power disparity among users leads to lower PAPR, enhancing power amplifier efficiency.

Additionally, Fig.~\ref{fig-papr-noma} shows that the PAPR gap between CIM-CPM-SS, IM-CPM-SS, and CIM remains relatively stable across different power allocations. This consistency implies that the PAPR improvements of IM-CPM-SS and CIM-CPM-SS over CIM are mainly independent of the specific power allocation strategy. In the two-user case, increasing the power disparity yields an improvement of approximately $0.5$ \si{dB} in PAPR performance. In the three-user case, this improvement grows to about $1$ \si{dB}. This trend indicates that PAPR increases more significantly with the number of users, which aligns with intuition: as more users are superimposed in the NOMA downlink, the likelihood of higher instantaneous signal peaks increases, thereby elevating the PAPR.

\subsection{BER of NOMA Downlink Transmission}
 \label{Section-V-C}
 
\begin{figure}[!t]
    \centering
    \includegraphics[width=0.95\linewidth]{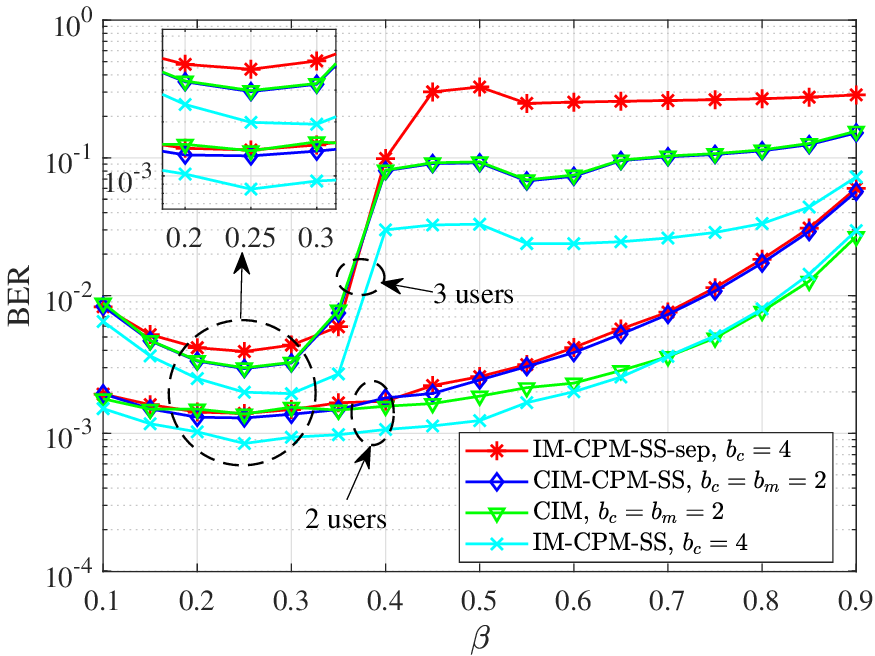}
    \caption{BER vs. power allocation factor ($\beta$) of CIM-CPM-SS, IM-CPM-SS, and CIM under NOMA scenario.}
    \label{fig-power-noma}
\end{figure}

Fig.~\ref{fig-power-noma} compares the BER performance of NOMA downlink transmissions for the CIM-CPM-SS, IM-CPM-SS, IM-CPM-SS-sep, and CIM schemes under various power allocation configurations. The analysis is performed at $E_b/N_0 = 30$ \si{dB}, and nonlinear distortion is not considered. The BER curves for all schemes exhibit a convex shape: initially decreasing, then increasing as the power allocation factor $\beta$ varies. This behavior results from the interplay between inter-user interference (IUI) and additive noise. When the power difference among users is small (i.e., $\beta$ is large), IUI dominates, resulting in a degraded BER. Conversely, when the power disparity is too significant (i.e., $\beta$ is small), the weaker users suffer from significant noise degradation. Thus, there exists an optimal power allocation that minimizes BER: $\beta = 0.25$ (with $P_1 = 0.8$ and $P_2 = 0.2$) for the two-user case, and $\beta = 0.25$ (with $P_1 = 0.7619$, $P_2 = 0.1905$, and $P_3 = 0.0476$) for the three-user case. These optimal values will be adopted in the subsequent simulation experiments.

\begin{figure}[!t]
    \centering
    \includegraphics[width=0.95\linewidth]{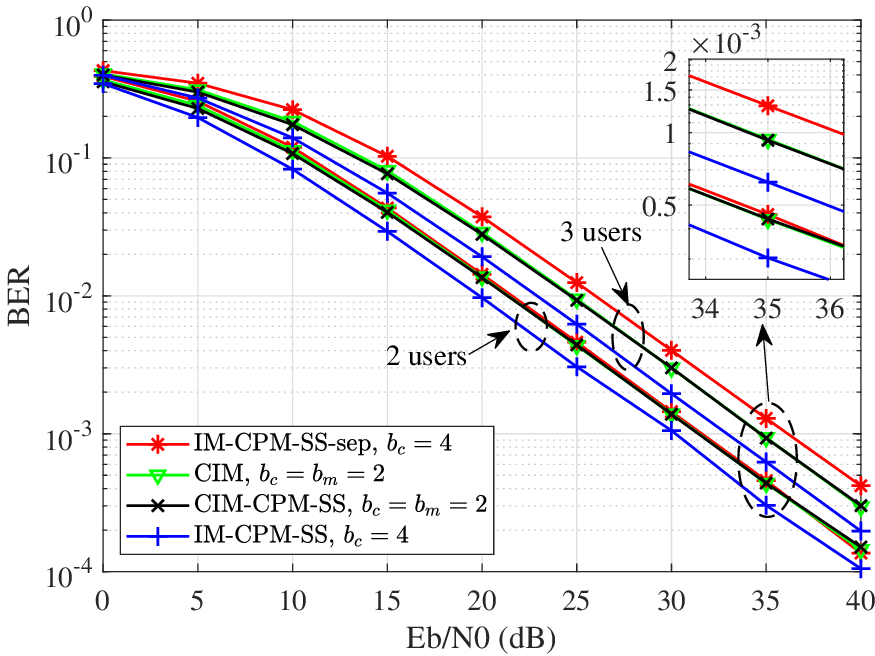}
    \caption{BER vs. $E_b/N_0$ of NOMA transmission for CIM-CPM-SS, IM-CPM-SS and CIM without nonlinear distortions when $\beta=0.25$.}
    \label{fig-ber-noma}
\end{figure}

Fig.~\ref{fig-ber-noma} presents the BER performance of the same schemes under varying $E_b/N_0$, again without nonlinear distortion. In the two-user case, the IM-CPM-SS scheme demonstrates superior performance, outperforming the CIM-CPM-SS and CIM schemes by approximately $1.5$ \si{dB} at a BER of $10^{-3}$. The CIM-CPM-SS, CIM, and IM-CPM-SS-sep schemes exhibit comparable BER performance. This outcome can be attributed to energy allocation per bit: at the same $E_b/N_0$ and spectral efficiency, each symbol of the IM-CPM-SS and IM-CPM-SS-sep schemes carries $4$ bits, while each symbol of the CIM-CPM-SS and CIM schemes carries only $2$ bits. Thus, the latter allocates more energy per bit, leading to a BER performance trade-off. Notably, the IM-CPM-SS-sep scheme still underperforms its joint-detection counterpart, as discussed earlier.

In the three-user case, the BER of the IM-CPM-SS-sep scheme further deteriorates compared to CIM and CIM-CPM-SS, highlighting the advantages of the proposed joint-detection schemes under more complex user interference conditions. This further confirms the effectiveness of IM-CPM-SS and CIM-CPM-SS in multi-user scenarios.

\begin{figure}[!t]
    \centering
    \includegraphics[width=0.95\linewidth]{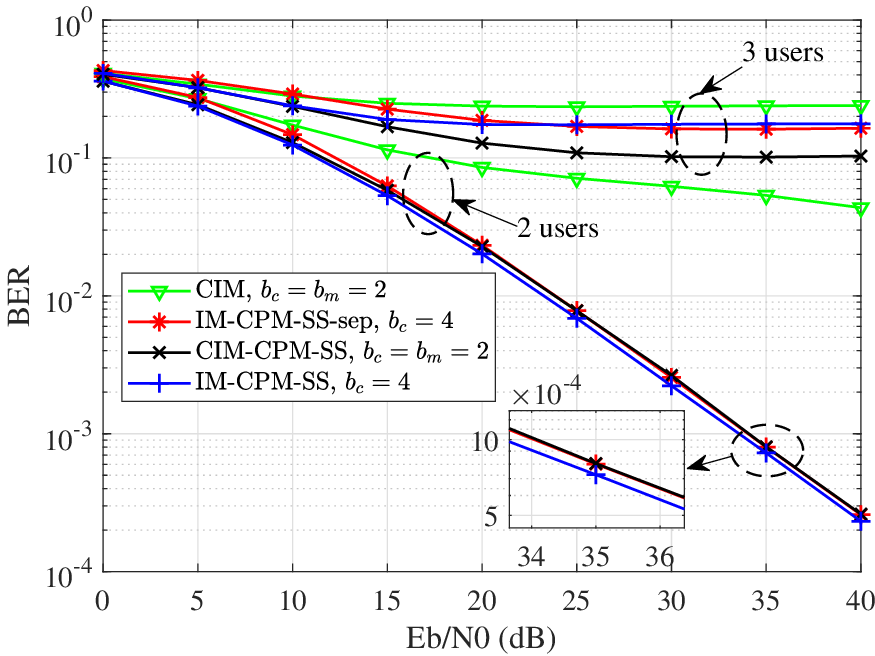}
    \caption{BER vs. $E_b/N_0$ of NOMA transmission for CIM-CPM-SS, IM-CPM-SS and CIM in the presence of nonlinear distortions when $\beta=0.25$.}
    \label{fig-rapp-noma}
\end{figure}

Fig.~\ref{fig-rapp-noma} investigates the impact of nonlinear distortion by evaluating BER under different IBO levels. The IBO values are determined based on PAPR measurements from Fig.~\ref{fig-papr-noma}. Specifically, in the two-user case, IBO is set to $2.55$ \si{dB} for IM-CPM-SS and IM-CPM-SS-sep, $3.03$ \si{dB} for CIM-CPM-SS, and $6.20$ \si{dB} for CIM. In the three-user scenario, the IBO values are $3.60$ \si{dB}, $4.21$ \si{dB}, and $7.30$ \si{dB} for the IM-CPM-SS (and IM-CPM-SS-sep), CIM-CPM-SS, and CIM schemes, respectively. These IBO levels correspond to the amplitude gain settings used in prior work \cite{CPM-NOMA}.

The results show that the CIM scheme suffers the most significant performance degradation under nonlinear distortion, with substantial increases in BER compared to the linear case (cf. Fig.~\ref{fig-ber-noma}). In contrast, the CIM-CPM-SS, IM-CPM-SS, and IM-CPM-SS-sep schemes experience only modest performance degradation, demonstrating their resilience to nonlinear effects. This robustness stems from their lower PAPR, making them more compatible with power-efficient transmitters and saturated power amplifiers.

However, in the three-user case, the effect of nonlinear distortion becomes more pronounced. With less power allocated to each user, the interference becomes more severe, resulting in noticeable BER degradation across all schemes. Nevertheless, even under such adverse conditions, CIM-CPM-SS and IM-CPM-SS maintain better performance than conventional CIM modulation, affirming their effectiveness in power-constrained, nonlinear multi-user communication environments. 

Notably, the proposed method demonstrates strong robustness against nonlinear distortions primarily due to the constant-envelope nature of the CPM signal. In LPWAN scenarios, low-cost and energy-efficient PAs are typically employed, which often operate in the nonlinear region to maximize power efficiency. These nonlinear PAs are highly sensitive to the envelope variations of the transmitted signal. A signal with large envelope fluctuations is more susceptible to nonlinear distortion when amplified, resulting in significant degradation of signal integrity. In contrast, CPM signals exhibit minimal envelope variation, allowing them to pass through nonlinear PAs with reduced distortion. As a result, the proposed scheme maintains signal fidelity even under nonlinear amplification.

\subsection{Advantages and Limitations of the Proposed Schemes}
The proposed IM-CPM-SS and CIM-CPM-SS schemes offer several notable advantages over conventional modulation techniques. First, by exploiting the quasi-orthogonality properties of CPM-SS sequences, both schemes significantly improve BER performance and data rate, thereby enabling more efficient spectrum utilization. This makes them particularly well-suited for bandwidth-constrained scenarios, such as massive machine-type communications in IoT networks. Second, both IM-CPM-SS and CIM-CPM-SS exhibit substantially lower PAPR compared to conventional CIM modulation. This low PAPR enhances robustness to nonlinear distortion introduced by power amplifiers, a critical factor in practical low-power and cost-sensitive IoT systems.

Despite these advantages, several trade-offs must be taken into account. The algorithmic complexity of both schemes, particularly in terms of joint detection and codeword correlation, may introduce significant computational overhead at the receiver. This can limit their feasibility in real-time or energy-constrained environments, such as battery-powered sensor nodes. Furthermore, the CIM-CPM-SS scheme relies on PSK modulation for conveying additional bits, which increases its sensitivity to phase noise and carrier frequency offset, especially in scenarios lacking tight synchronization. Additionally, while both schemes demonstrate strong performance in static or quasi-static channels, their robustness in dynamic environments characterized by mobility, Doppler spread, or co-channel interference requires further investigation. These challenges highlight the need for adaptive receiver designs and robust synchronization techniques to extend the applicability of the proposed schemes.

In summary, IM-CPM-SS and CIM-CPM-SS offer a compelling framework for enhancing downlink NOMA transmissions, combining improved BER and spectral efficiency with robust resilience to nonlinearities. Future research efforts should focus on reducing complexity, improving synchronization robustness, and validating performance in more diverse and dynamic channel conditions to facilitate real-world deployment in next-generation wireless systems.

\section{Conclusion}
 \label{Section-VI}
This paper proposed two novel modulation schemes, IM-CPM-SS and CIM-CPM-SS, that integrate IM or CIM with CPM-SS signaling for both point-to-point and NOMA downlink transmissions. By leveraging the quasi-orthogonality properties of CPM-SS waveforms, both schemes supported joint demodulation and despreading, which significantly enhances BER performance and spectral efficiency over conventional CPM-SS approaches. The IM-CPM-SS scheme increased the number of bits per symbol through code-domain indexing, while the CIM-CPM-SS scheme achieved even higher spectral efficiency by combining code-domain and symbol-domain indexing via PSK modulation. This enabled a flexible trade-off between error performance and throughput. Furthermore, both schemes exhibited lower PAPR than traditional CIM, thereby improving robustness to nonlinear distortion caused by power amplifier saturation—an important consideration in power-constrained IoT deployments. Simulation results demonstrated that both proposed schemes outperformed conventional separate detection methods in terms of BER and energy efficiency, particularly in NOMA scenarios. Moreover, their constant-envelope, continuous-phase characteristics maked them highly compatible with efficient radio front-ends. Overall, the proposed modulation schemes enabled constant-envelope, continuous-phase transmission with enhanced spectral efficiency, BER, and power efficiency. These characteristics maked them strong candidates for energy-efficient, high-reliability, and high-throughput IoT applications.

\begin{appendix}[Proof of Theorem~\ref{The-1}]
We analyze the cross-correlation coefficient between two CPM-SS symbols $\bm{y}_n$ and $\bm{y}_l$. By definition, the normalized cross-correlation is given by
\begin{align}
    \rho = \frac{\bm{y}_n^H \bm{y}_l}{\Vert \bm{y}_n \Vert_2 \, \Vert \bm{y}_l \Vert_2}.
\label{eq-crosscorr}
\end{align}
To evaluate $\rho$, we consider the specific case where the modulation index $h = 1/2$ and the memory length $L = 1$. Let $\bm{a}$ and $\bm{b}$ be the original $m$-sequence and its cyclically shifted version, respectively. Substituting these into \eqref{eq-CPM-v} and \eqref{eq-crosscorr}, we obtain
\begin{align}
    \rho &= \frac{1}{NP} \sum_{n=1}^{N} \sum_{k=1}^{P} \exp\left(j\left[ \pi q(k)(a_n - b_n) + \frac{\pi}{2} \sum_{i=1}^{n-1}(a_i - b_i) \right]\right),
\label{eq-crosscorr-v}
\end{align}
where $N$ is the code length and $P$ is the number of pulse samples.

Next, we simplify the expression of $\rho$ step by step. 
\begin{enumerate}[fullwidth, itemindent=1em, listparindent=1em]
    \item We begin counting the match and mismatch positions of two $m$-sequences. For a bipolar $m$-sequence $\bm{a}$ of length $N$, let $\bm{b}$ be the version obtained by a cyclic shift of $d$ positions. Then, the number of matching positions (i.e., $a_n - b_n = 0$) is $(N - 1)/2$, while the number of mismatching positions (i.e., $a_n - b_n = \pm 2$) is $(N + 1)/2$. Due to the balance property of $m$-sequences, mismatches are evenly distributed, and the counts of $+2$ and $-2$ are equal.
    \item Considering the inner summation for different $a_n - b_n$, we have
     \begin{align}
   \lefteqn{\sum_{k=1}^{P} \exp\left(j\pi q(k)(a_n - b_n)\right)} \nonumber \\
    &=
        \left\{\begin{array}{rl}
            P, & \text{if } a_n - b_n = 0, \\
            \sum\limits_{k=1}^{P} \exp(j 2\pi q(k)), & \text{if } a_n - b_n = 2, \\
            \sum\limits_{k=1}^{P} \exp(-j 2\pi q(k)), & \text{if } a_n - b_n = -2.
    \end{array}\right.
    \end{align}
    For randomly distributed mismatches, the contributions of $+2$ and $-2$ tend to cancel due to symmetry.  When $a_n - b_n = 0$, the contribution from matching positions simplifies to:
    \begin{align}
        T_0 = P \sum_{n: a_n = b_n} \exp\left(j \frac{\pi}{2} \sum_{i=1}^{n-1} (a_i - b_i)\right) = \pm P,
    \end{align}
    where the sign depends on the shift value $d$.

\item When mismatches occur, if the cyclic shift $d = \pm1$, the mismatches alternate as $\{+2, -2, +2, -2, \cdots\}$ or vice versa. In this case, the mismatch terms do not fully cancel out due to the regular alternating distribution of matches and mismatches. Instead, they contribute an additional imaginary component corresponding to $(N+1)/2$ terms. 

In contrast, if $d \neq \pm1$, the mismatches are not strictly interlaced, and the pseudo-random nature of the $m$-sequence ensures that the mismatches are pseudo-random. As a result, the contributions from these mismatches are fully canceled out due to the run properties of $m$-sequences, and no additional imaginary component appears in the cross-correlation result. Therefore, the cross-correlation coefficient $\rho$ can be approximated as 
\begin{equation}
    \rho = 
    \left\{\begin{array}{rl}
            \pm \frac{1}{N}, & \text {if } d \neq \pm1;     \\
            \pm \frac{1}{N} \pm j\frac{N+1}{2 NP} \sum\limits_{k=1}^P \sin(2 \pi q(k)),   & \text {if } d = \pm1. 
    \end{array}\right.
\label{eq-crosscorr-simp}
\end{equation}

When the frequency pulse REC is applied and $d = \pm1$, the cross-correlation coefficient becomes 
\begin{subequations}
\begin{align}
   \rho &= \pm \frac{1}{N} \pm j\frac{N+1}{2NP}\sum\limits_{k=1}^{P}\sin\left(\frac{\pi k}{P}\right) \label{eq-crosscorr-rec-a} \\
  	  &= \pm \frac{1}{N} \pm j\frac{N+1}{2NP} \cot\left(\frac{\pi}{2P}\right) \label{eq-crosscorr-rec-b} \\
	  & \approx \pm\frac{1}{N} \pm j\frac{N+1}{N\pi} \label{eq-crosscorr-rec-c}  \\
	  & \approx \pm\frac{1}{N} \pm j\frac{1}{\pi}, \label{eq-crosscorr-rec} 
\end{align}
\end{subequations}
where \eqref{eq-crosscorr-rec-c} is due to the approximation $\cot(\pi/2P) \approx 2P/\pi$ for large $P$. Therefore, as $N \to \infty$, \eqref{eq-crosscorr-simp} together with \eqref{eq-crosscorr-rec} reduces to the desired \eqref{eq-crosscorr-final}. This completes the proof of Theorem~\ref{The-1}.
\end{enumerate}
\end{appendix}

\bibliographystyle{IEEEtran}
\bibliography{MyRef}

\begin{IEEEbiography}
	[{\includegraphics[width=1in, height=1.25in, clip, keepaspectratio]{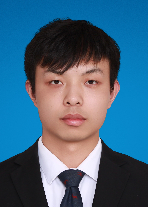}}]{Long Yuan} received the B.Eng degree from the School of Information Science \& Engineering, Lanzhou University, Lanzhou, China, in 2023. He is currently pursuing the M.Eng. degree at the School of Information Technology, Sun Yat-sen University, Guangzhou, China. His research interests include index modulation (IM), non-orthogonal multiple access (NOMA), multiple-input multiple-output (MIMO), the Internet of Things (IoT), and signal processing.
\end{IEEEbiography}

\begin{IEEEbiography}
	[{\includegraphics[width=1in, height=1.25in, clip, keepaspectratio]{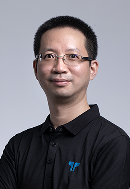}}]{Wenkun Wen} (Member, IEEE) received the Ph.D. degree in Telecommunications and Information Systems from Sun Yat-sen University, Guangzhou, China, in 2007. Since 2020, he has been with Techphant Technologies Co. Ltd., Guangzhou, China, as Chief Engineer.

From 2008 to 2009, he was with the Guangdong-Nortel R\&D center in Guangzhou, China, where he worked as a system engineer for 4G systems. From 2009 to 2012, he worked at the LTE R\&D center of New Postcom Equipment Co. Ltd., Guangzhou, China, where he served as the 4G standard team manager. From 2012 to 2018, he was with the 7th Institute of China Electronic Technology Corporation (CETC) as an expert in wireless communications. From 2018 to 2020, he served as Deputy Director of the 5G Innovation Center at CETC. His research interests include 5G/B5G mobile communications, machine-type communications, narrow-band wireless communications, and signal processing.
\end{IEEEbiography}	

\vfill

\begin{IEEEbiography}
	[{\includegraphics[width=1in, height=1.25in, clip, keepaspectratio]{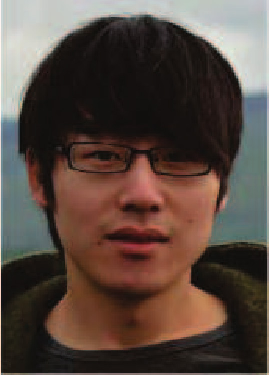}}]{Peiran Wu} (Member, IEEE) received the Ph.D. degree in electrical and computer engineering from The University of British Columbia (UBC), Vancouver, Canada, in 2015.
	
	From October 2015 to December 2016, he was a Post-Doctoral Fellow at UBC. In the Summer of 2014, he was a Visiting Scholar with the Institute for Digital Communications, Friedrich-Alexander-University Erlangen-Nuremberg (FAU), Erlangen, Germany. Since February 2017, he has been with Sun Yat-sen University, Guangzhou, China, where he is currently an Associate Professor. Since 2019, he has been an Adjunct Associate Professor with the Southern Marine Science and Engineering Guangdong Laboratory, Zhuhai, China. His research interests include mobile edge computing, wireless power transfer, and energy-efficient wireless communications. 
\end{IEEEbiography}

\vfill

\begin{IEEEbiography}
	[{\includegraphics[width=1in, height=1.25in, clip, keepaspectratio]{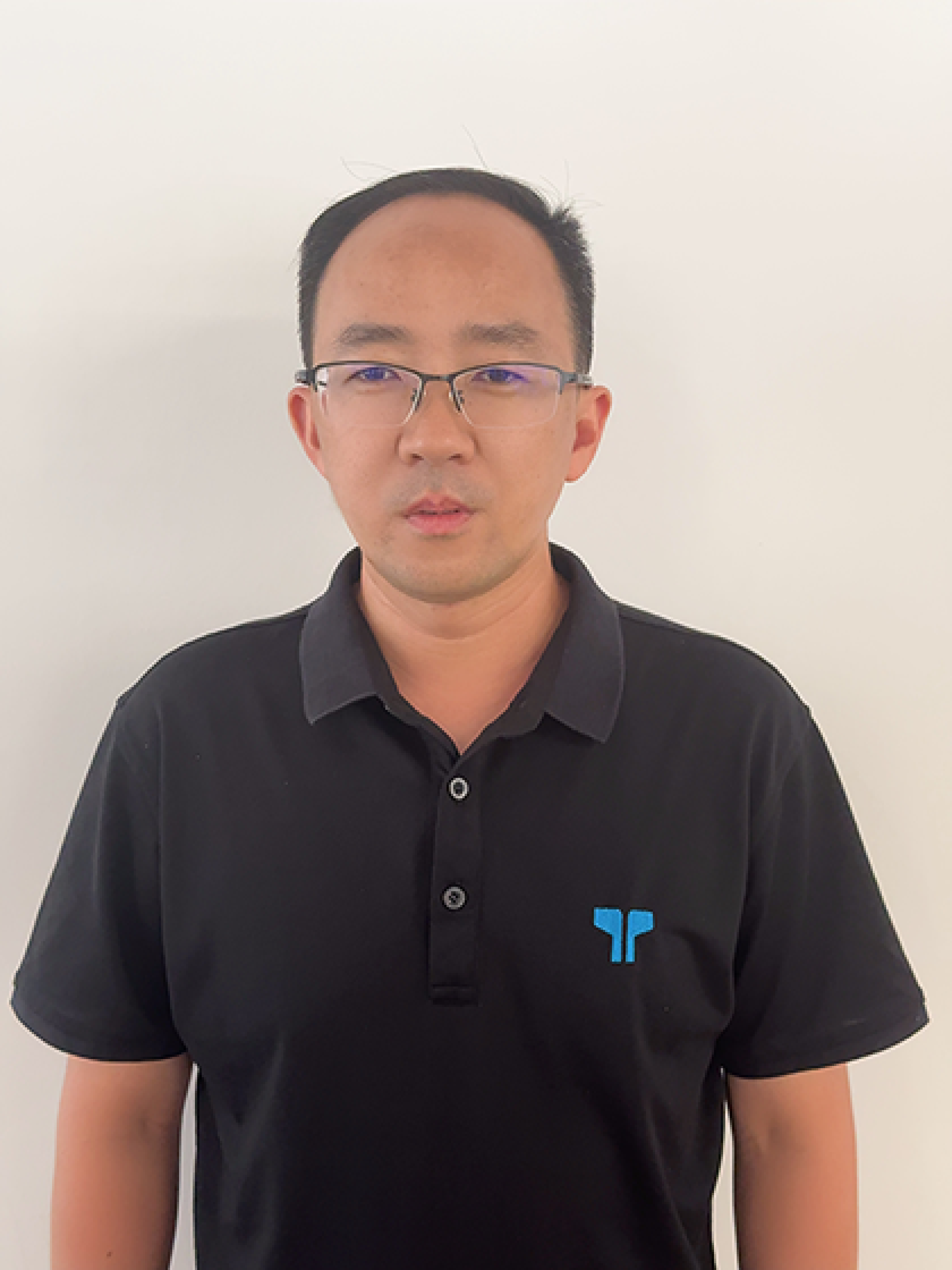}}]{Junlin Liu} received the M.S. degree in Telecommunications and Information Systems from Sun Yat-sen University, Guangzhou, China, in 2005. Since 2019, he has been with Techphant Technologies Co. Ltd., Guangzhou, China, as a system engineer. 
	
	From 2005 to 2009, he was with the Technology Center of the Guangzhou Jiesai Technologies Co. Ltd., Guangzhou, China, where he worked as a system engineer for WiMAX systems. From 2009 to 2012, he worked at the LTE R\&D center of New Postcom Equipment Co. Ltd., Guangzhou, China, where he served as a system engineer for 4G systems. From 2012 to 2019, he was with the 7th Institute of China Electronic Technology Corporation (CETC) as a system engineer in wireless communications. His research interests include 5G/B5G mobile communications, machine-type communications, narrow-band wireless communications, and signal processing.
\end{IEEEbiography}

\vfill
	
\begin{IEEEbiography}
	[{\includegraphics[width=1in, height=1.25in, clip, keepaspectratio]{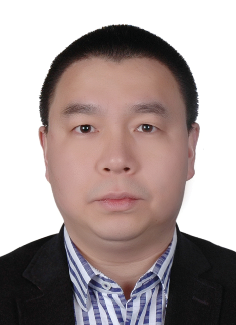}}]{Minghua Xia} (Senior Member, IEEE) received the Ph.D. degree in Telecommunications and Information Systems from Sun Yat-sen University, Guangzhou, China, in 2007.
	
	From 2007 to 2009, he was with the Electronics and Telecommunications Research Institute (ETRI) of South Korea, Beijing R\&D Center, Beijing, China, where he worked as a member and then as a senior member of the engineering staff. From 2010 to 2014, he was in sequence with The University of Hong Kong, Hong Kong, China; King Abdullah University of Science and Technology, Jeddah, Saudi Arabia; and the Institut National de la Recherche Scientifique (INRS), University of Quebec, Montreal, Canada, as a Postdoctoral Fellow. Since 2015, he has been a Professor at Sun Yat-sen University. Since 2019, he has also been an Adjunct Professor with the Southern Marine Science and Engineering Guangdong Laboratory (Zhuhai). His research interests are in the general areas of wireless communications and signal processing.
\end{IEEEbiography}

\vfill

\end{document}